\documentclass[iop]{emulateapj}
\usepackage{graphicx}

\begin{document}
\newcommand{\vdag}{(v)^\dagger}
\newcommand{\myemail}{skywalker@galaxy.far.far.away}

\shorttitle{No overdensity of Lyman Alpha Emitting Galaxies around a quasar at $\lowercase{z}\sim$5.7}
\shortauthors{Mazzucchelli et al.}

\title{No overdensity of Lyman Alpha Emitting Galaxies\\
   around a quasar at $\lowercase{z}\sim$5.7
    }

\author{C. Mazzucchelli\altaffilmark{1},
E. Ba{\~n}ados\altaffilmark{1,2,$\dagger$},
R. Decarli\altaffilmark{1},
E.P. Farina\altaffilmark{1},
B.P. Venemans\altaffilmark{1},
F. Walter\altaffilmark{1},
R. Overzier\altaffilmark{3}}
\affil{\altaffilmark{1}Max Planck Institute for Astronomy, K{\"o}nigstuhl 17, D-69117 Heidelberg, Germany}
\affil{\altaffilmark{2}The Observatories of the Carnegie Institute of Washington, 813 Santa Barbara Street, Pasadena, CA 91101, USA}
%\affil{\altaffilmark{3}Department of Astronomy, University of Texas at Austin, 1 University Station C1400, Austin, TX 78712, USA}
\affil{\altaffilmark{3}Observat$\rm \acute{o}$rio Nacional, Rua Jos$\rm \acute{e}$ Cristino, 77. CEP 20921-400, S$\mathrm{\tilde{a}}$o Crist$\rm \acute{o}$v$\mathrm{\tilde{a}}$o, Rio de Janeiro-RJ, Brazil}
\affil{\altaffilmark{$\dagger$}Carnegie-Princeton Fellow}
%\email{mazzucchelli@mpia.de}

\begin{abstract}
Bright quasars, observed when the Universe was less than one billion years old ($z>$5.5), are known to host massive black holes ($\sim$10$^{9}$ M$_{\odot}$), and are thought to reside in the center of massive dark matter overdensities. In this picture, overdensities of galaxies are expected around high redshift quasars. 
However, observations based on the detection of Lyman Break Galaxies (LBGs) around these quasars do not offer a clear picture: this may be due to the uncertain redshift constraints of LBGs, which are selected through broad-band filters only.
To circumvent such uncertainties, we here perform a search for Lyman Alpha Emitting galaxies (LAEs) in the field of the quasar PSO J215.1512$-$16.0417 at $z\sim$5.73, through narrow band, deep imaging with FORS2 at the VLT. We study an area of 37 arcmin$^{2}$, i.e. $\sim$206 comoving Mpc$^{2}$ at the redshift of the quasar. 
We find no  evidence for an overdensity of LAEs in the quasar field with respect to blank field studies. Possible explanations for these findings include that our survey volume is too small, or that the strong ionizing radiation from the quasar hinders galaxy formation in its immediate proximity.
Another possibility is that these quasars are not situated in the dense environments predicted by some simulations.
\end{abstract}


\section{Introduction} \label{secIntro}
Quasars are among the most luminous sources in the Universe, hosting supermassive black holes (as massive as M$\rm _{BH}\gtrsim$ few $\times$ 10$^{8}$ M$_{\odot}$) that reside in the center of massive galaxies. Quasars have been observed at high redshift ($z\gtrsim$5.5, e.g. \citealt{Fan06}, \citealt{Willott10}, \citealt{Morganson12}, \citealt{Banados14}, \citeyear{Banados16}, \citealt{Carnall15})  up to $z\sim$7  (e.g. \citealt{Mortlock11}, \citealt{Venemans13}, \citeyear{Venemans15}).
Metallicity estimates based on the intensity of broad emission lines (e.g. \citealt{Barth03}, \citealt{DeRosa11}) and observations of dust and molecular gas reservoirs (e.g. \citealt{Walter03}, \citealt{Maiolino05}, \citealt{Riechers09}; see for a review \citealt{Carilli13}) suggest that the host galaxies of these quasars have already experienced sufficient star formation to pollute the interstellar medium with metals, up to close-to-solar abundances.
The presence of apparently evolved systems so early in the history of the Universe (i.e. $\sim$1 Gyr from the Big Bang) sets constraints on the current scenario of structure formation and galaxy evolution. Models of massive black hole formation in the early Universe invoke the direct collapse of a massive gaseous reservoir (e.g. \citealt{Haehnelt93}, \citealt{Begelman06}, \citealt{Latif15}), the creation of black hole seeds from the collapse of PopIII stars (e.g. \citealt{Bond84}), or the interplay between gas collapse, star formation and dynamical processes (e.g. \citealt{DevecchiVolonteri09}; see for a review \citealt{Volonteri10}). In all these cases, the process of gas cooling and fragmentation drives the formation of both massive black holes and stars. A coexistence of fast-accreting black holes and star forming galaxies is thus expected at these very early cosmic epochs. Also, mergers among these galaxies may funnel black hole growth and could explain the onset of the black hole -- host galaxy scaling relations (see, e.g. \citealt{Jahnke11}).
%%%%%%%%%%%%%
%%%%%%%%%%%%%%%%%%% REFEREE %%%%%%%%%%%%%%%%%%%%%%%
%%%%%%%%%%%%%
The subsequent relation observed between the latter and the hosting dark matter halo suggests that these high-$z$ quasars, harboring black holes of $\sim$10$^{9}$ M$_{\odot}$, are expected to reside in massive dark matter halos of $\sim10^{13}$ M$_{\odot}$ (e.g. \citealt{Ferrarese02}, \citealt{WyitheLoeb03}), where also a large number of galaxies are expected to form \citep{Overzier09}.
%%%%%%%%%%%%%
%%%%%%%%%%%%%%%%%%% REFEREE %%%%%%%%%%%%%%%%%%%%%%%
%%%%%%%%%%%%%
These structures can eventually evolve into large gravitationally bound systems in the present Universe (with a large scatter in mass, from groups to clusters, e.g. \citealt{Springel05}, \citealt{Overzier09}, \citealt{Angulo12}).
%%%%%%%%%%%%%
%%%%%%%%%%%%%%%%%%% REFEREE %%%%%%%%%%%%%%%%%%%%%%%
%%%%%%%%%%%%%
Observational attempts to detect these high-redshift galaxies in the vicinities of high redshift quasars complement the aforementioned theoretical predictions.
%%%%%%%%%%%%%
%%%%%%%%%%%%%%%%%%% REFEREE %%%%%%%%%%%%%%%%%%%%%%%
%%%%%%%%%%%%%
One of the most efficient and successful methods to identify high redshift galaxies is through the Lyman Break technique (drop-outs or Lyman Break Galaxies, LBGs): absorption by neutral hydrogen causes a break in the observed galactic spectrum, apparent from abrupt change in colors (e.g. \citealt{Steidel96}).\\
A number of studies investigated the presence of the theoretically expected galaxies around $z\sim$6 quasars, using the Lyman Break technique. However, the picture sketched out by observations is far from clear. For instance, \cite{Stiavelli05} found an overdensity of LBGs in the field of the bright $z\sim$6.28 quasar SDSS J1030+0524, based on $i_{775}$ and $z_{850}$ images taken with the Advanced Camera for Surveys (ACS) at the \textit{Hubble Space Telescope} (\textit{HST}, with a field of view of $\rm \sim11~arcmin^{2}$, corresponding to an area of $\sim$65 comoving Mpc$^2$ [cMpc$^2$] at $z\sim$6). On the other hand, \cite{Kim09} studied the environment of five $z\sim$5 quasars, again searching for LBGs through \textit{HST} ACS imaging: they estimated that the fields around two quasars are overdense, two underdense and one consistent with a blank field. \cite{Simpson14} investigated the environment of the most distant quasar to date, ULAS J1120+0641 ($z\sim$7), recovering no evidence for the presence of an overdensity of galaxies (using data from \textit{HST} ACS). Studies on scales larger than ACS \textit{HST} also do not provide an unambiguous scenario. \cite{Utsumi10} found an enhancement in the number of LBGs in the field of the quasar CFHQS J2329-0301 ($z\sim$6.4), observed with the Suprime Camera at the Subaru Telescope, whose field of view covers an area of $\sim$900 arcmin$^{2}(\sim$4600 cMpc$^{2}$). Recently,  \cite{Morselli14} showed that four $z\sim$6 quasars were situated in overdense environments, based on a search for LBGs with deep multi-wavelength photometry from the Large Binocular Camera (LBC) at the Large Binocular Telescope (whose field of view covers a wide area of $\rm \sim575~arcmin^{2}\sim$3100 cMpc$^{2}$).
Conversely, \cite{Willott05} imaged three SDSS quasars at $z>$6 with GMS-North on the Gemini-North Telescope (whose field of view is $\rm \sim30~arcmin^{2}$, amounting to $\sim$170 cMpc$^{2}$ at $z\sim$6), recovering no clear signs for an overdensity of LBGs.\\
The different findings may be ascribed to different reasons, e.g. the depths reached in the observations, diverse techniques/selection criteria considered, different survey areas (and therefore scales) probed, and the diverse fields inspected, which may be intrinsically different. More importantly, the Lyman Break technique, using broad-band filters whose pass-bands normally span $\rm \Delta \lambda\sim 1000~\mathrm{\AA}$, does not provide an accurate redshift determination ($\rm \Delta z \sim$1, equals to line of sight distances $\rm \Delta d \sim 850$ cMpc or $120$ physical Mpc [pMpc] at $z\sim$6). Any possible overdensity may be diluted over this big cosmological volume \citep{Chiang13}, taking into account that the Universe is homogeneous at scales $\gtrsim$ 70-100 $h^{-1}$ Mpc (\citealt{Wu99}, \citealt{Sarkar09}).
Samples of photometric LBGs, selected without a large number of broad band filters, are likely to be contaminated by foreground sources (e.g. lower-$z$, red/dusty galaxies).
 
A more secure approach to identify high redshift galaxies is to look for sources with a bright Ly$\alpha$ emission (Lyman Alpha Emitters, LAEs). Such narrow line emission can be recovered by specific narrow band filters ($\rm \Delta \lambda \sim100~\mathrm{\AA}$). An immediate advantage, with respect to the LBG selection, is that the redshift range covered is much narrower ($\Delta z \sim$ 0.1, corresponding to $\rm \Delta d \sim 44$ cMpc$\sim 7$ pMpc at $z\sim$6), i.e. an overdensity membership can be clearly established.
%%%%%%%%%%%%%
%%%%%%%%%%%%%%%%%%% REFEREE %%%%%%%%%%%%%%%%%%%%%%%
%%%%%%%%%%%%%
LAEs are believed to be mostly low-mass galaxies, spanning a range of stellar masses of $\sim$ 10$^{6}-$10$^{8}$ M$_{\odot}$ and ages of $\sim$1$-$3 Myr (e.g. \citealt{Pirzkal07}, \citealt{Ono10}).
However, a non negligible fraction of massive galaxies (with masses up to $\sim$10$^{11}$ M$_{\odot}$), and galaxies hosting older stellar population ($\sim$ 1 Gyr) has been also found among LAEs (e.g. \citealt{Pentericci09}, \citealt{FinkelsteinS09}, \citealt{FinkelsteinK15}).
%%%%%%%%%%%%%
%%%%%%%%%%%%%%%%%%% REFEREE %%%%%%%%%%%%%%%%%%%%%%%
%%%%%%%%%%%%%
\citeauthor{B13} (\citeyear{B13}, hereafter B13) carried out the first and, until the present work, only search for LAEs at scales $\gtrsim$1 pMpc around a $z>$5.5 quasar\footnote{Previously, also \cite{Decarli12} searched for Ly$\alpha$ emission around two $z>$6 quasars through narrow band imaging with \textit{HST}, but their study was limited to far smaller scales ($\sim$1 arcmin$\sim$0.35 pMpc).}. They used a collection of broad and narrow band filters at the Very Large Telescope (VLT). No strong evidence was found for an enhancement in the number of LAEs with respect to the blank field.
 
Here we present a search for LAEs in the field around the Broad Absorption Line quasar PSO J215.1512$-$16.0417 (hereafter PSO J215$-$16).
%The source is a Broad Absorption Lines (BAL) quasar, characterized by a blue continuum.
%%%%%%%%%%%%%
%%%%%%%%%%%%%%%%%%% REFEREE %%%%%%%%%%%%%%%%%%%%%%%
%%%%%%%%%%%%%
It has a bolometric luminosity of 3.8$\times$10$^{47}$ erg s$^{-1}$, with 0.2 dex of uncertainty, and a black hole mass of 6.7$\times$10$^{9}$ M$_{\odot}$, with an uncertainty of 0.3 dex. The redshift is $z$=5.732$\pm$0.007,
%0.007%,
measured from the OI emission line ($\lambda_{\mathrm{rest}}=1307$ \AA). This line is the brightest and clearest among the emission lines observed in the spectrum of the quasar. As a further check, other emission lines were fitted (N{\tiny V}, SII, CII).
%after the Ly$\alpha$ peak wavelength was fixed using the OI redshift.
The redshift estimates obtained are consistent (with a scatter of $\sim$0.02) within the astrophysical systematic uncertainties (\citealt{Morganson12}).
%n general, redshifts measured from broad emission lines should be adopted with caution: indeed these values may differ of $\sim$400 km s$^{-1}$ from the secure estimates obtained from [CII] or CO lines (\citealt{Venemans16}). Unfortunately, no radio or mm observations of PSO J215$-$16 are available to date.
%%%%%%%%%%%%%
%%%%%%%%%%%%%%%%%%% REFEREE %%%%%%%%%%%%%%%%%%%%%%%
%%%%%%%%%%%%%
The structure of this paper is as follows: we describe the observations and data reduction in Section 2. In Section 3 we present our LAEs selection criteria. In Section 4 we study the environment of the quasar on the base of the candidates found in the previous section.
%%%%%%%%%%%%%
%%%%%%%%%%%%%%%%%%% REFEREE %%%%%%%%%%%%%%%%%%%%%%%
%%%%%%%%%%%%%
In Section 5 we simulate a population of $z\sim$5.7 LAEs to which compare our results. In Section 6 we place our work in the context of the current clustering studies.
%%%%%%%%%%%%%
%%%%%%%%%%%%%%%%%%% REFEREE %%%%%%%%%%%%%%%%%%%%%%%
%%%%%%%%%%%%%
%In Section 6 we obtain an estimate of the mass of the dark matter host halo, based on the quasar properties, and we derive the number of LAEs expected around the quasar. 
In Section 7 we discuss our results. Finally, in Section 8 we present our conclusions and outline possible future steps.\\
All magnitudes reported here are in the AB system and corrected for the galactic dust extinction following \cite{Schlafly11} . We use a $\rm \Lambda$CDM cosmology, with $\rm H_{0} = 70$ km s$^{-1}$ Mpc$^{-1}$, $\rm \Omega_{M} = 0.3$, $\Omega_{\Lambda} = 0.7$: this implies that 1 arcmin $\sim$ 2.36 cMpc $\sim$ 0.35 pMpc at $z$=5.73 (the redshift of the quasar studied here).

\section{Observations and Data Reduction}
\begin{figure}
\epsscale{.80}
\centering
\includegraphics[width=0.8\columnwidth]{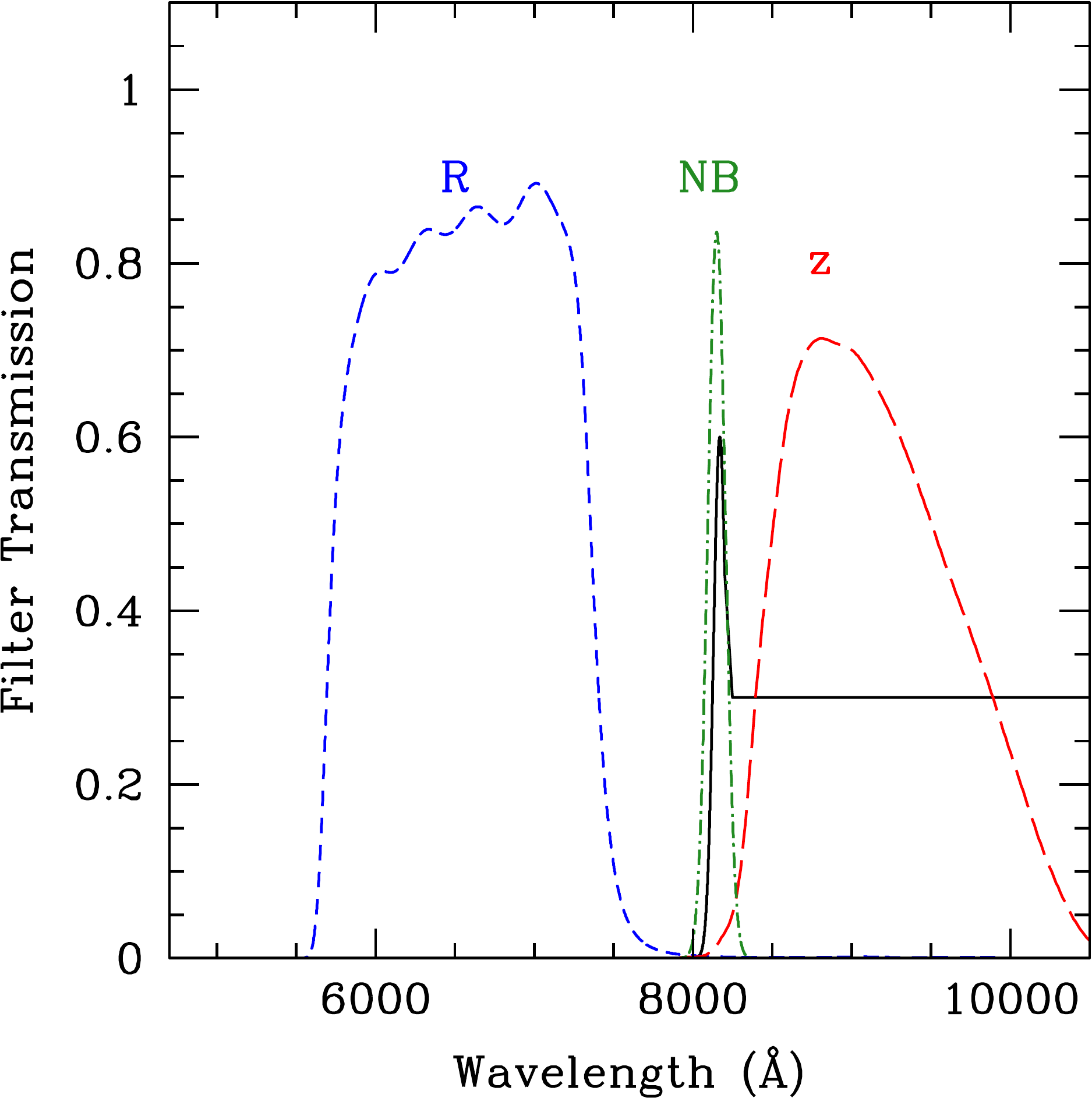}
\caption{Set of filters used in the present study, R\_SPECIAL (R, \textit{short-dashed blue line}), z\_GUNN (z, \textit{long-dashed red line}) and the Narrow Band filter FILT815\_13+70 (NB, \textit{dot-dashed green line}). In \textit{solid black line}, a synthetic spectrum of a Lyman Alpha Emitter (LAE) at the redshift of the quasar studied in this work ($z\sim$5.7).}
\label{figFilt}
\end{figure}

\begin{figure*}
\epsscale{.80}
\centering
\includegraphics[width=\textwidth]{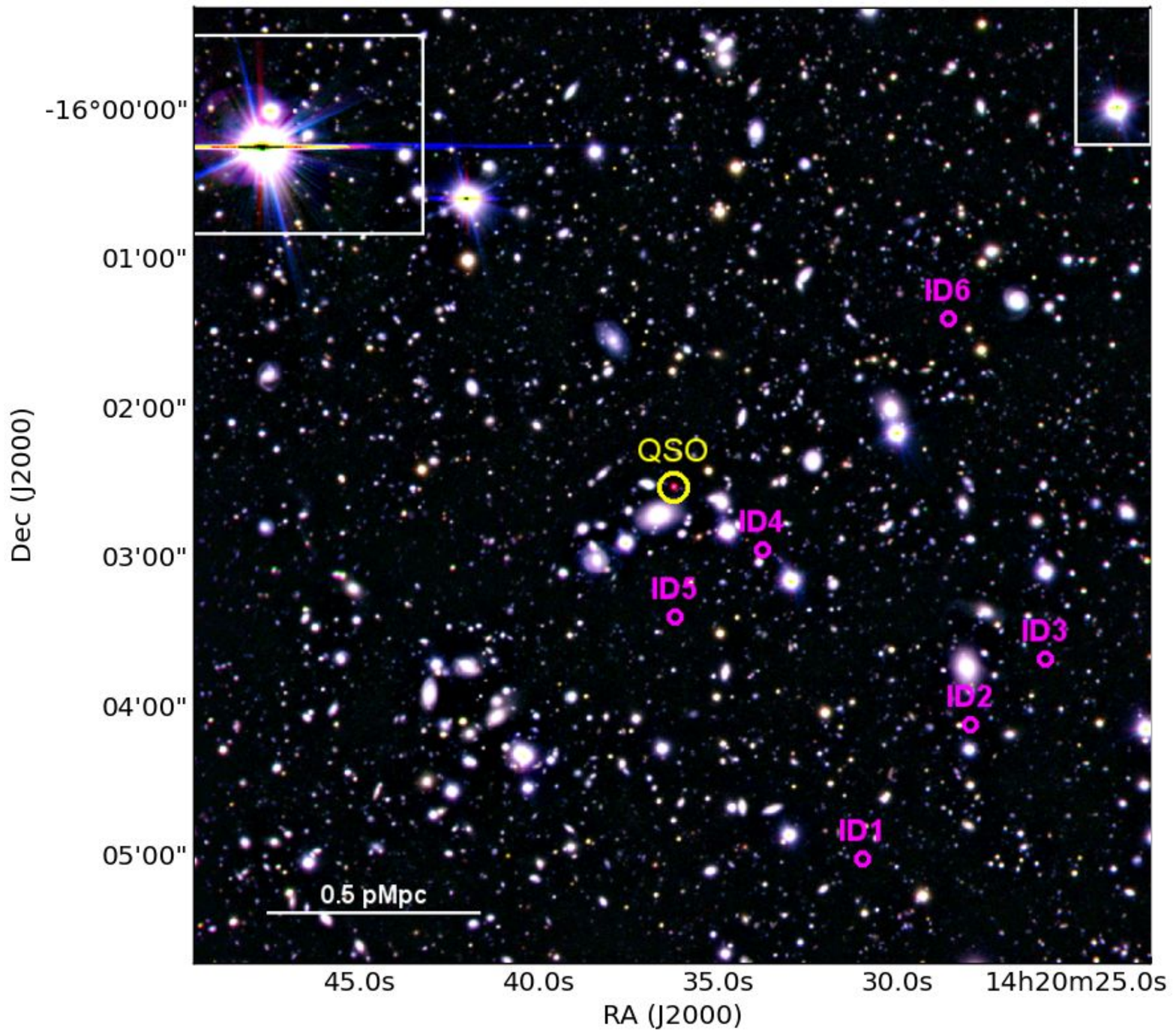}
\caption{RGB composite image of the field around the quasar PSOJ215$-$16. In \textit{magenta} we show the position of the sources only observed in the narrow band, but not detected (at 2$\sigma$ confidence level) in the R and z filters (see Section 3, Fig 3 and Table 1.). The position of the quasar and masked regions around bright stars are also shown. The total area analyzed is 37 arcmin$^{2}$.
\label{figRGB}}
\end{figure*}

We obtained multi-wavelength photometry of the field around the quasar PSO J215$-$16 with the FOcal Reducer/low dispersion Spectrograph 2 (FORS2, \citealt{app92}) at the VLT. The observations were obtained over nine nights in 2013, June, July and August. We used the red sensitive detector consisting of two 2k$\times$4k MIT CCDs. In order to decrease the read out time and noise, we adopted a 2$\times$2 binning. The resulting pixel size is 0.25 arcsec/pixel and the total field of view is equal to 6.8$\times$6.8 arcmin$^{2}$, i.e. 2.38$\times$2.38 pMpc$^{2}$.\\
We  collected images in two broad band filters R\_SPECIAL (R, with a central wavelength $\rm \lambda_{c}=6550~\mathrm{\AA}$, and a width $\rm \Delta\lambda=1650~\mathrm{\AA}$) and z\_GUNN (z, $\rm \lambda_{c}=9100~\mathrm{\AA}$,  $\rm \Delta\lambda=1305~\mathrm{\AA}$), and in the narrow band filter FILT815\_13+70 (NB, $\rm \lambda_{c}=8150~\mathrm{\AA}$, $\rm \Delta\lambda=130~\mathrm{\AA}$). The filters allow us to select LAEs at redshifts between $ 5.66 \lesssim z \lesssim 5.75$ ($ \mathrm{\Delta} z\sim0.1$), i.e. at the precise redshift of the quasar studied here. Using the broad filters, LBGs can be selected in a redshift range of $ 5.2 \lesssim z \lesssim 6.5$ ($\mathrm{\Delta} z\sim1.3$). The filter throughputs, together with a synthetic LAE spectrum at the redshift of the quasar, are shown in Fig. \ref{figFilt}.\\
%-------------------
%-------------------
The individual exposure times in each frame are 240\,s in R, 115 \,s in z and 770 \,s in NB. Each exposure was acquired with a dithering of $\sim$10$\arcsec$, in order to account for bad pixels and remove cosmic rays. The total exposure times are, respectively, $\sim$1.13, 2.11 and 8.56 hr in the R, z and NB filter. \\
We perform a standard data reduction: we subtract from each exposure the bias, we apply the flat field and we subtract the background, then the images are aligned and finally combined; the astrometric solution was derived with \texttt{astrometry.net} \citep{Lang}.
A composite RGB image of the quasar field is shown in Fig. \ref{figRGB}.\\
%---------------------
%---------------------
The seeing values of the stacked images are equal to 0.$\arcsec$78 in R, 0.$\arcsec$81 in z and 0.$\arcsec$79 in NB. In order to circumvent uncertainties due to different apertures or to the angular resolution of the images, we match the Point Spread Function (PSF) of the R and NB images to the one of the z filter frame (the one with the worst seeing), using the IRAF task gauss.
We calculate the photometry using, as reference sources, field stars retrieved from the Pan-STARRS1 catalog \citep{Magnier13}. We calculate the conversion between the two different filter sets by interpolating spectra of standard stars. The relations found are :
%---------------
\begin{eqnarray}
\rm R = r_{P1} - 0.277\times(r_{P1}-i_{P1}) -0.005\\
\rm  z  = z_{P1} -0.263\times(z_{P1}-y_{P1})-0.001\\
\rm  NB = i_{P1} -0.626\times(i_{P1}-z_{P1})+0.014
\end{eqnarray}
%--------------
where r$\rm _{P1}$, i$\rm _{P1}$, z$\rm _{P1}$ and y$\rm _{P1}$ are the magnitudes in the Pan-STARRS1 filters.\\
The obtained zero points values are 27.77 $\pm$ 0.04 in R, 27.12 $\pm$ 0.02 in z and 24.85 $\pm$ 0.04 in the NB filter.
%%%%%%%%%%%%%
%%%%%%%%%%%%%%%%%%% REFEREE %%%%%%%%%%%%%%%%%%%%%%%
%%%%%%%%%%%%%
%We calculate the noise in our images by estimating the value of the background RMS in a circular aperture of 3 pixels radius, by placing apertures of this size randomly in empty regions in the images. 
We calculate the noise in our images by computing the standard deviation of flux measurements in
circular apertures of 3 pixels radius that were placed randomly in empty regions in the
images.
This is the radius of the aperture over which we perform our photometry, as discussed below. We achieve limit magnitudes at 5$\sigma$ level of 26.47, 25.96 and 26.38 mag in the R, z and NB frames.
%%%%%%%%%%%%%
%%%%%%%%%%%%%%%%%%% REFEREE %%%%%%%%%%%%%%%%%%%%%%%
%%%%%%%%%%%%%

\begin{figure}
\epsscale{.80}
\centering
\includegraphics[width=0.8\columnwidth]{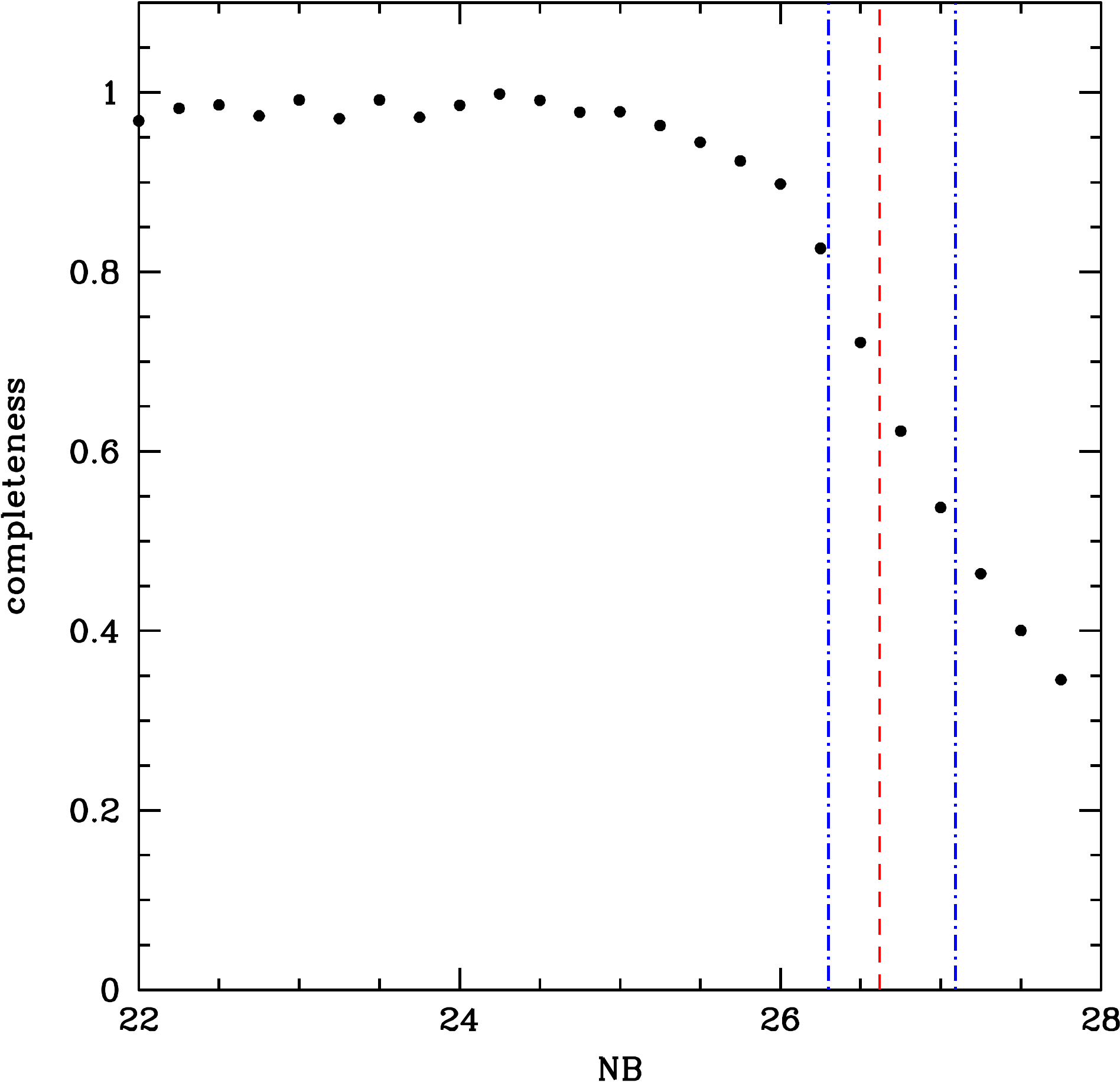}
\caption{Detection completeness function for the sources in our catalog, as a function of NB magnitude (see text for details). With a \textit{red, long-dashed line} we show the 4$\sigma$ limit magnitude estimated from the image noise, which corresponds to a level of completeness of 67\%. We show the 80\% and 50\% completeness levels (at magnitudes of 26.3 and 27.1, respectively), with \textit{blue, dot-dashed lines}.}
\label{figCompFun}
\end{figure}

In order to identify sources from the images, we use the software \texttt{SExtractor} \citep{Bertin} in the double image mode, requiring a minimum detection threshold of 1.8$\sigma$. Since we expect LAEs to be strongly detected in the NB filter, we take this frame as the base of our selection.
We cut the outskirts of the frames and mask saturated stars, where both the astrometry and the photometry were less reliable. The final effective area is equal to 37 arcmin$^{2}$ (i.e. $\sim$206 cMpc$^{2}$ at the redshift of the quasar).\\
%The total remaining sky area covered is equal to ~38.48 arcmin$^{2}$.\\
We perform the photometry of the sources over an aperture of 3 pixel radius (0.$\arcsec$75). For our frames this is equal to 1.8$\times$seeing: i.e. we encompass more the $\sim$90\% of the flux and, at the same time, maximize the signal-to-noise ratio.\\
We consider only sources with signal-to-noise ratio S/N$>$4 in the NB filter, and adopt a 2$\sigma$ upper limit in R and z for non-detections in the broad band images (27.46 in R and 26.95 in z). Sources non detected in the broad bands are allowed in the catalog, and we substitute the R and z values with the respective $2\sigma$ limit magnitudes.
Finally, we use the `flags' parameter given by \texttt{SExtractor} in order to discard unreliable detections, rejecting sources with flags $\geq$ 4 (i.e. objects saturated, truncated, or whose aperture data are incomplete or corrupted); the final catalog encompasses 3250 sources.\\
%%%%%%%%%%%%%
%%%%%%%%%%%%%%%%%%% REFEREE %%%%%%%%%%%%%%%%%%%%%%%
%%%%%%%%%%%%%
We extrapolate the cumulative count for the sources detected in our NB frame, in order to estimate the completeness function of our study at the faint end (see Fig. \ref{figCompFun}). We compute the logarithmic cumulative number counts of sources detected in NB as a function of NB magnitude. We fit it with a linear relation (in log-mag space) for 21$<$NB$<$25, and extrapolate it towards the faint end. The completeness is computed as the ratio between the expected number counts from the logN-logS extrapolation and the actual number of detected sources. Our catalog reaches a completeness of 80\% and 50\% at NB magnitudes of 26.3 and 27.1, respectively.  

%-------------------
%---------------------

%A simple extrapolation of the logN$-$logS relation towards faint sources suggests that our catalog is complete to 90\% level for objects with NB $<$ 26.2. 

\section{Selection of High Redshift Galaxy Candidates} \label{secSel}
In this work we follow the color selection defined in B13, and briefly described hereafter.

LAEs are expected to be well detected in the narrow band and to show a break in the continuum emission. More precisely, we required our LAE candidates to satisfy the following criteria:
%----------------
\begin{itemize}
\item (z$-$NB) $>$ 0.75\\
We request the flux density in the NB to be at least twice the one observed in the z filter. This cut implies that we are selecting objects with an Equivalent Width (EW) of the Ly$\alpha$ line in the rest frame greater than 25.$~\mathrm{\AA}$ (see Section \ref{secSim}).
\item (R$-$z) $>$ 1.0\\
We expect lower flux at wavelength shorter than the Ly$\alpha$ emission line, i.e. a break in the spectrum, which can be identified by requiring a very red R-z color. In case a source is not detected in R or z, we adopt the 2$\sigma$ limit magnitudes.
\item $\rm |(z-NB)|>2.5 \times \sqrt{\sigma_{z}^{2}+\sigma_{NB}^{2}}$\\
We want to select only objects with a significant flux excess in the NB. Therefore, we adopt a constrain in order to discard all the objects satisfying our selection criteria only due to their photometric errors.
\item NB $>$ 18\\
Since LAEs are expected to be faint sources at these redshifts, we impose a lower limit to the observed NB magnitude. However, we note that there are no objects with NB$<$18 that satisfy all the previous criteria.
\end{itemize}
%-------------------
\begin{figure}
\epsscale{.80}
\centering
\includegraphics[width=\columnwidth]{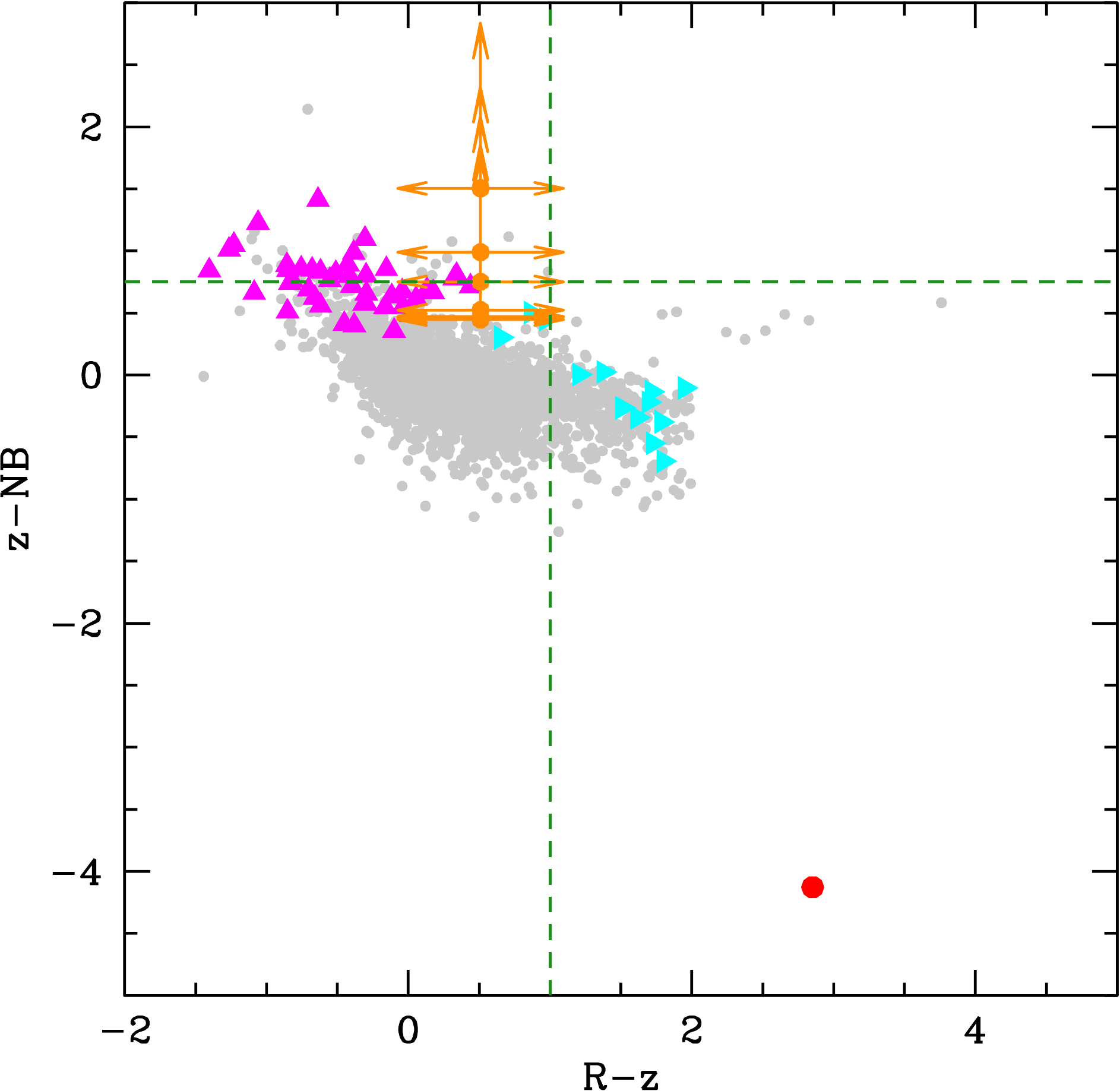}
\caption{Color-Color diagram of the sources in our field detected in the NB filter with SNR$>$4. The sources detected with a significance lower than 2$\sigma$, only in the z or in the R filter, are shown as \textit{magenta} and \textit{cyan triangles} respectively. \textit{Orange arrows} indicate objects undetected in both R and z (6 sources). Our selection cuts for LAEs are displayed with \textit{green dashed lines}; LAEs should fall in the upper right panel (see Section 3 for the complete set of our criteria). No LAE candidates are securely found in our field. The \textit{red point} in the lower right corner corresponds to the quasar.}
\label{figColorColor}
\end{figure}
%---------------------
%----------------      
In Fig. \ref{figColorColor} we show the (z$-$NB) vs (R$-$z) color-color diagram, together with our high redshift galaxies selection. LAEs are expected to fall in the upper right part.
In summary, we find no secure detections of LAEs in our field, i.e. no sources fully satisfy all the selection criteria described above.
We observe six sources with a detection in the NB, that are not detected in both R and z frame. The (z$\rm _{2\sigma,lim}$-NB) color ranges from a value of 0.45 to 1.5. 
In the following analysis, we conservatively consider all these six sources as LAE candidates; however, we stress that, in order to know if these objects would fully satisfy our criteria, deeper R and z band observations are needed.
\\
In Table \ref{tab1} we report their coordinates, the NB magnitudes, the projected distances from the quasar, the estimated Ly$\alpha$ luminosities and the Star Formation Rates (which are within the expectations for typical $z\sim6$ LAEs, SFR$\sim$6$^{+3}_{-2}$ M$_{\odot}$ yr$^{-1}$, \citealt{Ouchi08}; see Appendix \ref{AppendixA}). Their postage stamps in the three filters are shown in Fig. \ref{figPS}.

%-------------------
\begin{deluxetable*}{ccccccccc}
%\begin{deluxetable}
%\begin{center}
%\centering
\tabletypesize{\small}
\tablecaption{Source names, sky Coordinates, NB magnitudes (AB system) and projected distances to the quasar of the objects retrieved in our field with a detection in NB (at S/N$>$4) and a non--detection in the broad bands. Also, we show the luminosities of the Ly$\rm \alpha$ emission line and the star formation rates (SFRs) as estimated in Appendix \ref{secSFR}. The errors on the SFRs are derived from the photometric uncertainties on the narrow band magnitudes, and do not account for systematics in the underlying assumptions .\label{tab1}}
\tablewidth{0pt}
\tablehead{
\colhead{ID} & \colhead{RA} & \colhead{DEC} & \colhead{mag$\rm _{NB}$} & \colhead{Angular} & \colhead{Comoving} & \colhead{Physical} & \colhead{$L_{Ly\alpha}$} &  \colhead{SFR} \\
   & & & & \colhead{Distance} & \colhead{Distance} & \colhead{Distance} & &  \\
   & \colhead{(J2000.00)} & \colhead{(J2000.00)} & \colhead{[AB]} & \colhead{[arcmin]} & \colhead{[cMpc]} & \colhead{[pMpc]} & \colhead{$\times$10$^{42}$ [erg s$^{-1}$]} & \colhead{[M$\rm _{\odot}$ yr$^{-1}$]} 
}
\startdata
ID1 & 14:20:31.1 & -16:04:59.2 & 26.50 $\pm$ 0.22 & 2.79 & 6.58 & 0.98 & $>$ 1.9 $\pm$ 0.4 & 1.2 $\pm$ 0.2\\
ID2 & 14:20:28.1 & -16:04:05.9 & 25.45 $\pm$ 0.09 & 2.55 & 6.00 & 0.89 & $>$ 5.1 $\pm$ 0.4 & 3.2 $\pm$ 0.3\\
ID3 & 14:20:26.0 & -16:03:39.7 & 26.43 $\pm$ 0.21 & 2.74 & 6.46 & 0.96 & $>$ 2.1 $\pm$ 0.4 & 1.3 $\pm$ 0.2\\
ID4 & 14:20:33.9 & -16:02:55.9 & 25.96 $\pm$ 0.14 & 0.74 & 1.75 & 0.26 & $>$ 3.2 $\pm$ 0.4 & 2.0 $\pm$ 0.2\\
ID5 & 14:20:36.3 & -16;03:23.0 & 26.20 $\pm$ 0.17 & 0.88 & 2.08 & 0.31 & $>$ 2.5 $\pm$ 0.4 & 1.6 $\pm$ 0.2\\
ID6 & 14:20:28.7 & -16.01:23.8 & 26.48 $\pm$ 0.22 & 2.14 & 5.05 & 0.75 & $>$ 2.0 $\pm$ 0.4 & 1.2 $\pm$ 0.2
\enddata
\end{deluxetable*}
%---------------------
\begin{figure}
\epsscale{.80}
\centering
\includegraphics[width=\columnwidth]{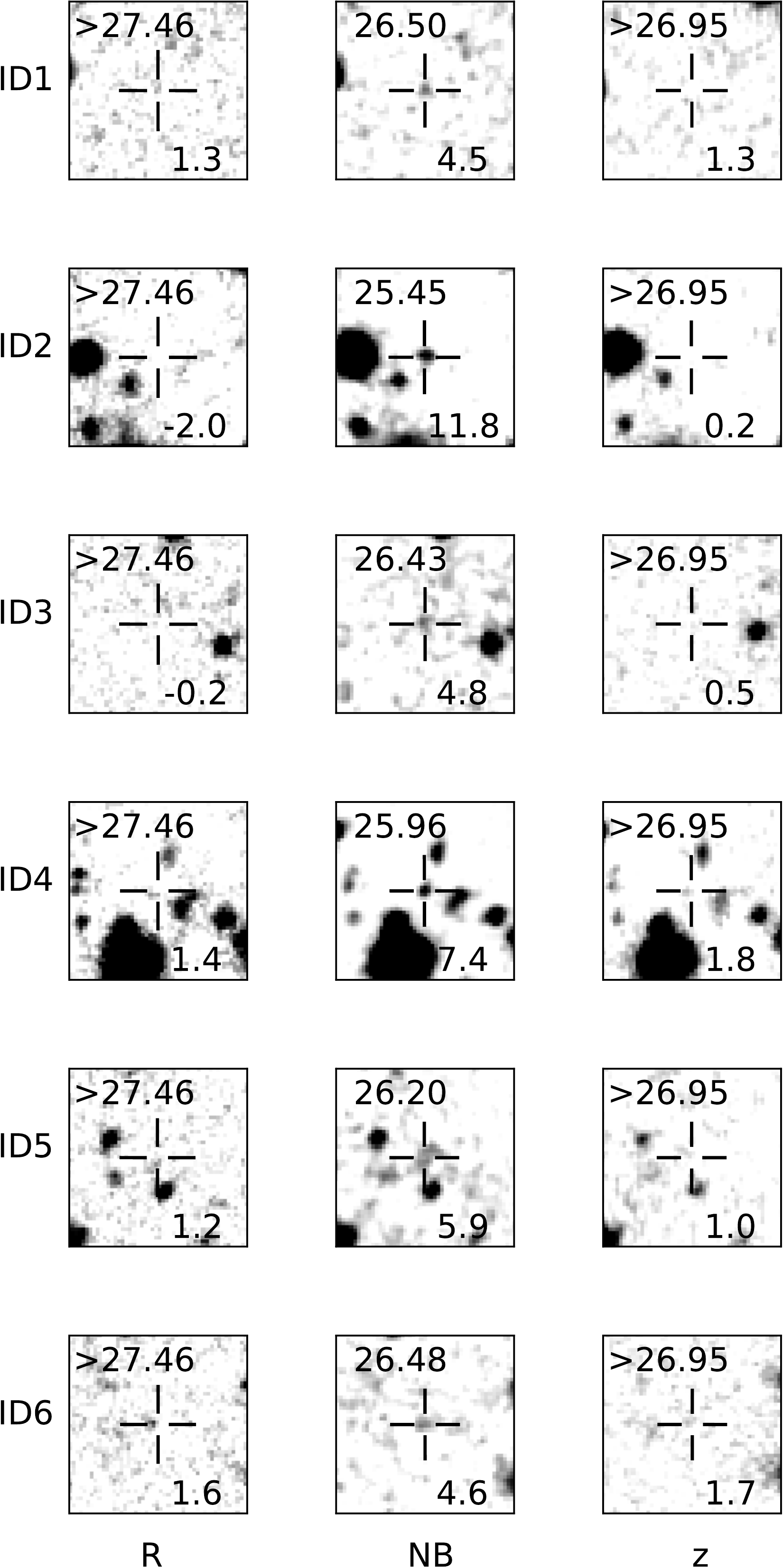}
\caption{Postage Stamps centered on the sources detected only in the NB, in a region of 12$\arcsec\times$12$\arcsec$. The magnitudes and S/N in the three bands are also reported in top left and bottom right corner, respectively.}
\label{figPS}
\end{figure}
%---------------------
\section{Study of the Environment} \label{secNumCount}
\begin{figure}
\epsscale{.80}
\centering
\includegraphics[width=\columnwidth]{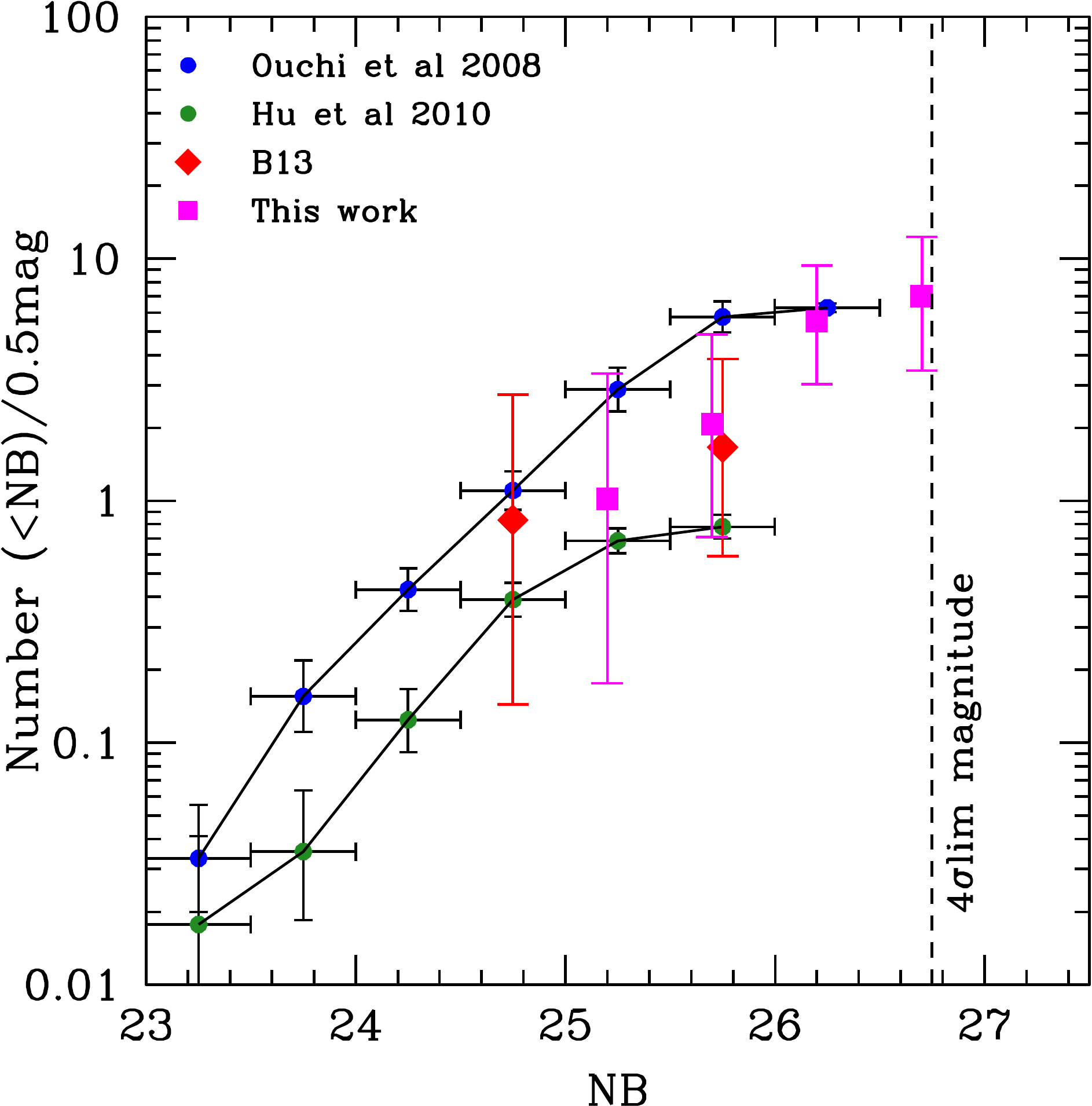}
\caption{Cumulative number counts of LAEs observed in two blank fields, i.e. not containing quasars, by \cite{Ouchi08} and \cite{Hu10} (\textit{blue} and \textit{green dots}) and in the field of the $z$=5.7 quasar ULAS J0203+0012 by B13 (\textit{red diamonds}). All the number counts are re-scaled to our effective area. Also, we show the 4$\sigma$ limit magnitude in this work (\textit{dashed line}). In the present study we retrieve six possible LAE candidates (\textit{magenta squares}, see Section 3). The number counts of this work are corrected taking into account our completeness at the corresponding NB magnitudes. We have no evidence for an overdensity around the quasar. The errors reported are the poissonian noise for small counts (\citealt{Gehrels86}).}
\label{figNumbCount}
\end{figure}

In order to study the environment of the quasar PSO J215$-$16, we compare our findings both to earlier quasar environment studies, and to blank fields (i.e. fields where no quasars are present).
In Fig. \ref{figNumbCount} we show the cumulative number counts, rescaled to our effective area (37 arcmin$^{2}$), of LAEs found in two blank fields, \cite{Ouchi08} and \cite{Hu10}, and in the field of another $z\sim$5.7 quasar, B13. The number counts of the objects found in this work are corrected taking into account the completeness of our catalog at the respective NB magnitude.
%---------------------
%---------------------

\cite{Ouchi08} and \cite{Hu10} searched for LAEs at redshift $z=$5.7 in seven SuprimeCam fields and in the Subaru/XMM-Newton Deep Survey (SXDS), respectively. The total areas covered in the two studies are $\sim$1.16 and $\sim$1 deg$^{2}$. The difference between the number counts of these two measurements may be ascribed to the fact that \cite{Hu10} consider only the spectroscopically confirmed sources in their sample, while the sample of \cite{Ouchi08} is based on the photometric selection, possibly affected by contaminants but also characterized by a higher completeness level.
The present work and B13 are carried out with the same filter set and instrument, and assuming consistent selection criteria.
Even if we consider all the sources detected solely in the NB filter in our work as LAE candidates (see Section 3), the number counts would be consistent with the blank field measurements and with B13.

Therefore, there is no evidence for an overdensity of LAEs at the redshift of the quasar.

\section{Simulation of LAE} \label{secSim}
\begin{figure}
\epsscale{.80}
\centering
\includegraphics[width=\columnwidth]{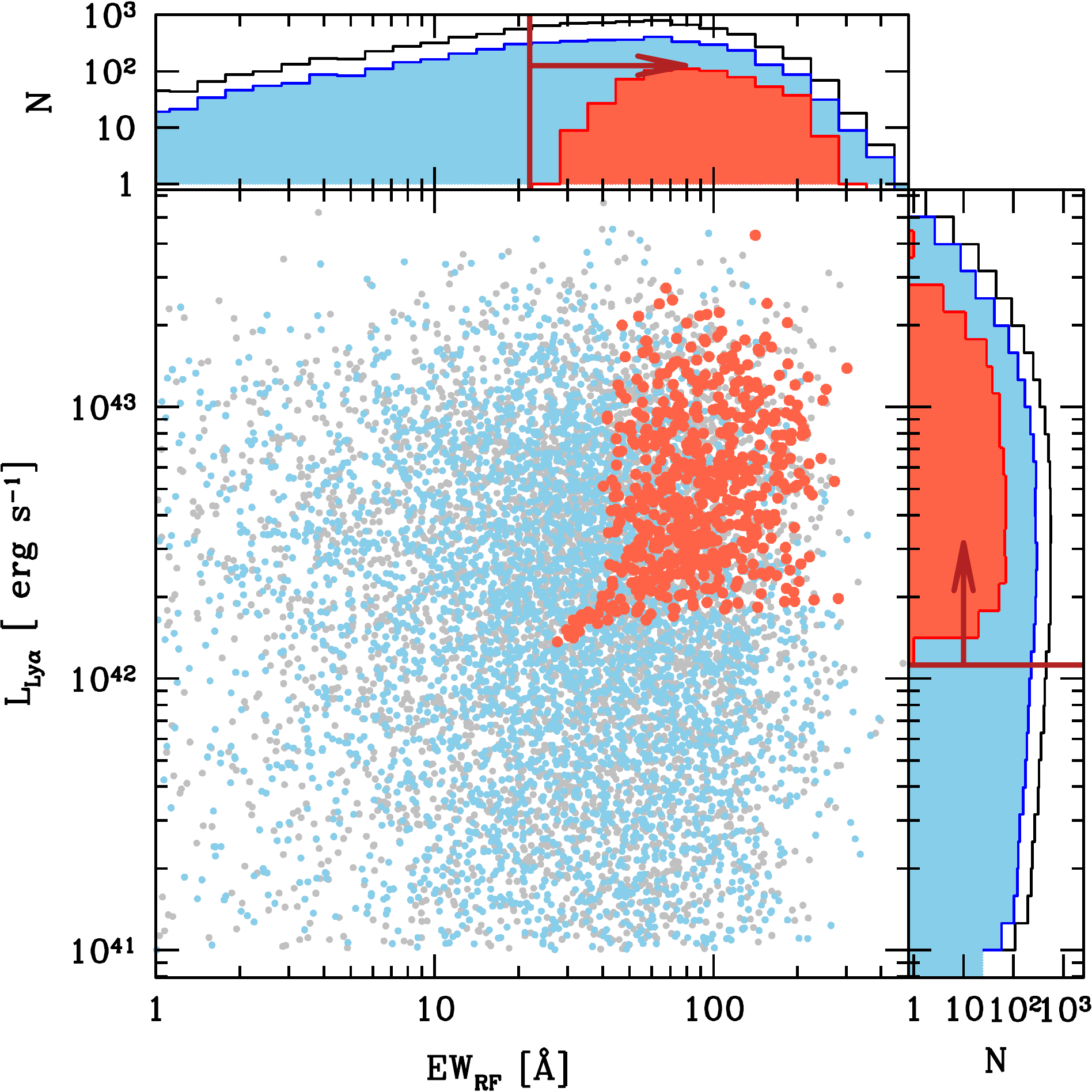}
\caption{Equivalent width at rest-frame and Ly$\alpha$ luminosity of all our synthetic templates of LAEs (\textit{grey}), of sources in the  redshift range 5.61$<$z$<$5.77 (\textit{blue}), and of the ones detected as LAEs by our selection criteria (\textit{red points}). The \textit{upper} and \textit{right} panels show the distribution of EW and Ly$\alpha$ luminosity, respectively. The \textit{red, solid lines} shows the minimum EW and Ly$\alpha$ luminosity retrieved (25 \AA\ and 1.32$\times$10$^{42}$ erg s$^{-1}$, respectively).}
\label{figHistoDetLAE}
\end{figure}

\begin{figure}
\epsscale{.80}
\centering
\includegraphics[width=\columnwidth]{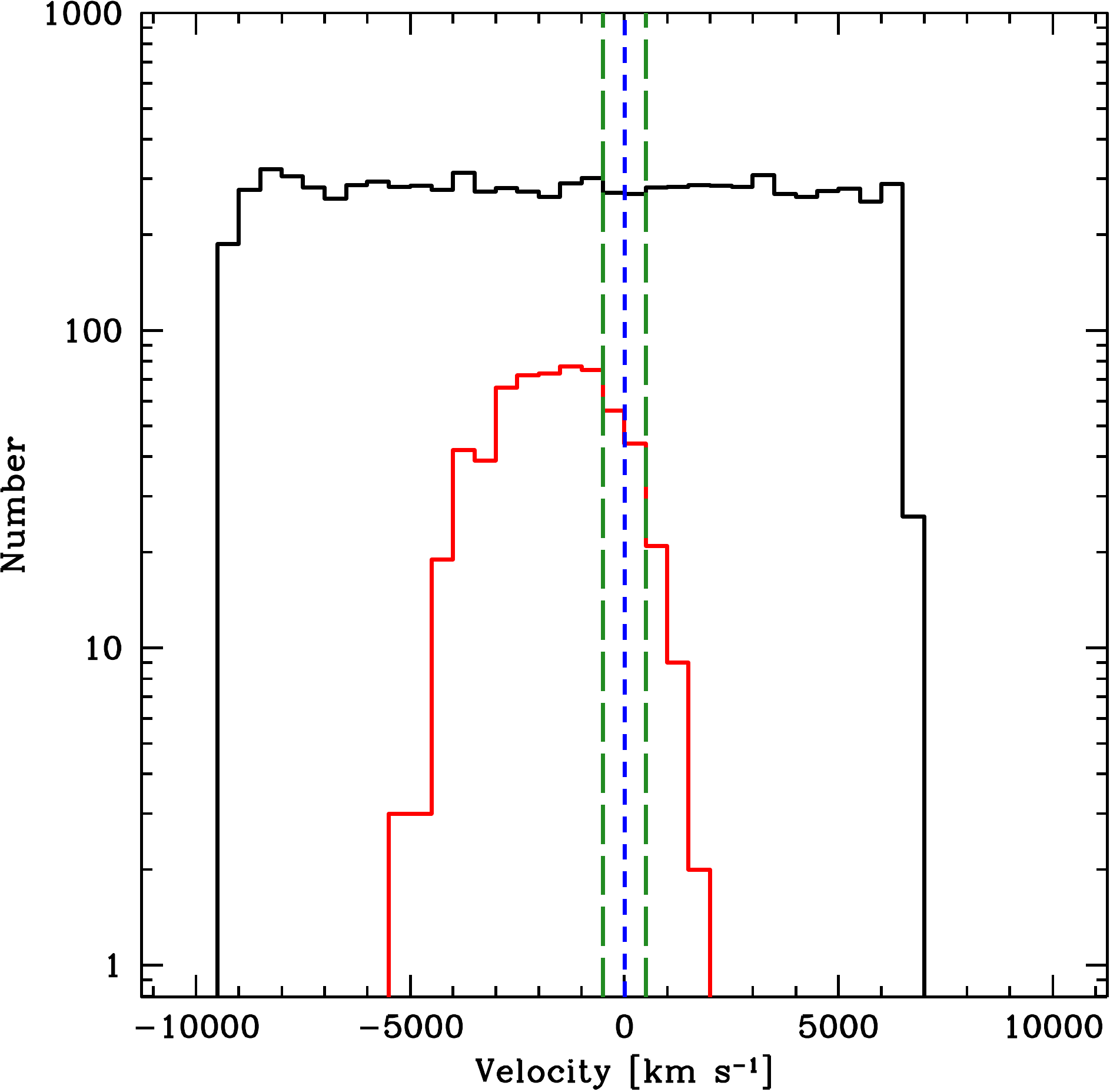}
\caption{Velocity distribution of mocked LAEs, with respect to the systemic velocity of PSO J215$-$16 (\textit{short-dashed, blue line}). In \textit{black} we report the distribution of all the simulated sources, while in \textit{red} only the sources recovered as LAEs. The velocity window encompassing +500 km s$^{-1}$/-500 km s$^{-1}$ with respect to the quasar is shown with \textit{long-dashed, green lines}. The velocity window is not centered on the redshift of the quasar, but we still recover $\gtrsim$50\% of the expected galaxies population gravitationally bound to the quasar.}
\label{figVelDist}
\end{figure}

%%%%%%%%%%%%%
%%%%%%%%%%%%%%%%%%% REFEREE %%%%%%%%%%%%%%%%%%%%%%%
%%%%%%%%%%%%%
We estimate the fraction of LAEs that we expect to detect, based on observations of blank fields, our filters set, and our image depths.
%We aim to assess if, given the set-up of our observations, we can truthfully sample a population of galaxies gravitationally bound to the quasar.
We aim to assess, given the set-up of our observations, our ability to recover a population of galaxies that is gravitationally bound to the quasar.
\\
We adopt a distribution of LAEs according to the luminosity function reported by \citeauthor{Ouchi08} (\citeyear{Ouchi08}, with parameters L*=6.8 $\times$10$^{42}$ erg s$^{-1}$, $\Phi$* = 7.7$\times$10$^{-4}$ Mpc$^{-3}$ and $\alpha$=-1.5).
%We first crudely estimate the number of sources we expect to detect in all our cosmological volume ($\sim$8264 Mpc$^{-3}$, where we considered the redshift range to be 5.66 $<$ z $<$ 5.75) by integrating the luminosity function till the luminosity limit allowed by the depth of our NB image (L$_{\mathrm{Ly\alpha, lim}}$=1.72 $\times$ 10$^{42}$ erg s$^{-1}$). We obtain an expected number of $\sim$9 sources.\\   
We then create a synthetic population of LAEs, drawing objects from the luminosity function through the Monte-Carlo method.
We modeled the LAEs as composed by a flat, continuum emission (L$_{\mathrm{cont}}$ = L$_{\mathrm{Ly\alpha}}$/EW) and a Ly$\alpha$ emission line, implemented as a Gaussian function, with FWHM at rest-frame of 200 km s$^{-1}$ (corresponding to typical line widths for LAEs, e.g., \citealt{Ouchi08}). We account for the absorption due to the intergalactic medium using the reshift-dependent recipe given by \cite{Meiksin06},
and assume an exponential distribution of equivalent widths at rest frame (\citealt{Zheng14}, N = $\rm e^{-EW/ 50 \mathrm{\AA}}$). 
Our mock spectra are randomly distributed in the redshift range $5.52<z<5.88$ and cover down to a luminosity of L$\rm _{Ly\alpha}$ = 10$^{41}$ erg s$^{-1}$.\\
%The final catalog counts 8989 objects.\\
This results in synthetic magnitudes in the filters used in this work. We consider the sources detected with our NB image depth, at 4$\sigma$ magnitude limit. We substitute the z and R broad band magnitudes with their respective 2$\sigma$ limit magnitudes, in case their values were fainter than our  detection limits.
%We report in Fig. \ref{figColorColorSynt} the (z-NB) vs (R-z) color-color plot, analogous to Fig. \ref{figColorColor}.\\
%---------------------
%\begin{figure}
%\epsscale{.80}
%\centering
%\includegraphics[width=\columnwidth]{cc_LAE_test9_c.pdf}
%\caption{Color-color (z-NB) vs (R-z) plot from synthetic LAEs template. We show the complete sample (down to a luminosity of the Ly$\alpha$ line of 10$^{41}$ erg s$^{-1}$) in \textit{grey points}, the sources detected with our NB image depth in \textit{black circle points}, the objects undetected solely in R, and both in R and z bands (down to a 2$\sigma$ significance in the broad bands), with \textit{blue triangles} and \textit{orange squares}, respectively. Our LAEs selection cuts are depicted in \textit{green dashed lines}.}
%\label{figColorColorSynt}
%\end{figure}
%---------------------
We considered the synthetic LAEs detected by our color selection criteria (see Section \ref{secSel}), and all the sources not detected in both the broad bands but detected in NB.\\
%We recover 756 sources.\\
In Fig. \ref{figHistoDetLAE}, we show the EW at rest-frame and Ly$\alpha$ luminosity distributions of all our generated LAEs, of the subsample in the redshift range 5.6$<z<$5.78 (the window in which we find LAEs in our simulation), and of LAEs recovered by the criteria presented in this study. We recover sources with a minimum EW at rest frame of 25 \AA, and a Ly$\alpha$ luminosity of 1.32$\times$10$^{42}$ erg s$^{-1}$. The last value is in agreement with what obtained if one calculates the limit L$_{Ly\alpha}$ by considering that all the flux observed in the NB filter, at our 4$\sigma$ limit, is due the line emission ($\sim$1.72$\times$10$^{42}$ erg s$^{-1}$, see Appendix \ref{secSFR}). \\
%---------------------
%---------------------
We normalize the total number of simulated sources to the number of objects expected in a blank field, in a cosmological volume equal to the one analyzed here, where we use a line-of-sight depth of 40 cMpc (see Section \ref{secIntro}). Integrating the LAE luminosity function down to the Ly$\alpha$ luminosity limit considered in the simulation (10$^{41}$erg s$^{-1}$) we expect to measure 81.4 sources.
Taking into account our depth and selection criteria, we recover 28\% of the original sample.
Therefore, we expect to observe $\sim$23 LAEs in a field of the same cosmological volume as our study. We actually selected six LAEs candidates in our images; correcting for a completeness level of 67\% at our NB magnitude limit (see Fig. \ref{figCompFun}), we obtain 8 objects. This comparison suggests that the field surrounding PSO J215$-$16 might be less than $\sim$3 times less dense than the blank field, in agreement with what is shown in Fig. \ref{figNumbCount}\footnote{We note that \cite{Ouchi08} states that their completeness level at their last luminosity bin (NB = 26) is estimated to be 50\%$-$60\%. Therefore, the last point of the blue curve in Fig \ref{figNumbCount} should be corrected accordingly.}.\\
We report in Fig. \ref{figVelDist} the logarithmic distribution of the LAEs systemic velocities with respect to the quasars, again, for all our simulated sample and for the objects detected as LAEs. The quasar is not located at the center of the redshift range covered by our selection. Nevertheless we can still recover a significant fraction of galaxies even at the red edge. If we assume that LAEs that are gravitationally bound to the quasar are distributed following a Gaussian function centered on the quasar systemic velocity and a width of $\sigma\sim500$ km s$^{-1}$, we calculate that, given the shifted position of the quasar, we can recover 56\% of the total LAE population. If we assume a velocity dispersion of $\sim$1000 km s$^{-1}$, our estimate decreases to 53\%. Several studies (e.g. \citealt{Hashimoto13}, \citealt{Song14}) suggest that the Ly$\alpha$ emission could be redshifted with respect to the systemic velocity of the source by $\sim$200 km s$^{-1}$. If this would be the case, we would be affected even more. Nevertheless, we find that the fraction of observed objects in this case would decrease only to 49\% and 48\% (in case of $\sigma$=500 km s$^{-1}$ and 1000 km s$^{-1}$, respectively). Thus, even in the most extreme scenario, we can still recover $\gtrsim$50\% of the expected LAEs present in the proximity of PSO J215$-$16.   
%%%%%%%%%%%%%
%%%%%%%%%%%%%%%%%%% REFEREE %%%%%%%%%%%%%%%%%%%%%%%
%%%%%%%%%%%%%
However, it is necessary to consider that our observed counts are also affected by Poisson noise and cosmic variance. \cite{Trenti08} provide estimates on the variance of the observed number counts, taking into account both Poisson noise and cosmic variance, and based on a variety of parameters (e.g. the survey volume and completeness, the halo filling factor and expected number of objects)\footnote{\url{http://casa.colorado.edu/~trenti/CosmicVariance.html}}. In our case, we consider a cosmological volume V given by our field of view and a redshift interval of $\Delta$z$\sim$0.16, centered on $z$=5.69 (V$\sim$14900 cMpc$^{3}$, see Fig. 8); an intrinsic number of objects equals to the one recovered by our LAE simulation (23), a completeness of 67\% (see Section 2 and Fig. \ref{figCompFun}) and a Sheth-Tormen bias calculation. \cite{Kovac07}, studying the clustering properties of LAEs at $z\sim$4.5, find a value for the duty cycle of high redshift LAEs varying within 6\%$-$50\%.
%\footnote{This estimate is however based on the clustering study of a lower redshift population ($z=$4.5).}
We consider here both extreme cases. However, \cite{Kovac07} consider a sample of LAEs with a minimum EW of 80 \AA; since our EW limit is lower (25 \AA, see Fig. \ref{figHistoDetLAE}), we expect that the halo filling factor for our case would be closer to 50\%.
% an intermediate value of 25\%.
Then, we expect to observe 15 $\pm$ 7 (9) sources in the case of a duty cycle of 6\% (50\%), where the fractional uncertainty due to cosmic variance and Poisson noise are 36\% (49\%) and 26\% (in both cases), respectively. These results are consistent within 1.3$\sigma$, in the first case, and 1$\sigma$ in the second case with what expected in the present study.
%Then, we expect to observe 15 $\pm$ 9 sources, where the fractional uncertainty due to cosmic variance and Poisson noise is 50\% and 25\%, respectively. This result is consistent within 1$\sigma$ with what observed in the present study.
Conversely, we can also obtain a rough estimate of the cosmic variance, for a population of galaxies for which we know the expected number density in a certain volume, following \cite{Somerville04}. They use cold dark matter models to derive the expected bias ($b$) and root variance of the dark matter ($\sigma_{\mathrm{DM}}$) as a function of galaxy number density and survey volume, respectively (see their Fig. 3). We consider our survey volume as reported above, and an expected number density obtained integrating the LAE luminosity function until the Ly$\alpha$ luminosity limit from our simulation ($\sim$10$^{-3}$ cMpc$^{-3}$); we derive a fractional cosmic variance of $\sigma_{\mathrm{v}}=b\sigma_{\mathrm{DM}}\sim$0.66 for a population of galaxies at z=6. This value is higher than what recovered using the method illustrated by \cite{Trenti08}; however the cosmic variance is not a trivial quantity to estimate, and depend on several assumptions considered in the models, e.g. the LAEs halo filling factor. Considering the latter value obtained, we would expect to detect 23 $\pm$ 15 sources, which, taking into account our completeness, is consistent within 1$\sigma$ with the observations reported here. Also, considering a clustering scenario, e.g. consistent with the LAE-LAE clustering case (see \citealt{Ouchi03} and Section \ref{secClust}), we would obtain 26 $\pm$ 18 sources, consistent with the observed ones.
%\footnote{We note that the latter estimate of the cosmic variance is higher than the value recovered using the method illustrated by \cite{Trenti08}. However, the cosmic variance is not a trivial quantity to estimate, and depend on several assumptions considered in the models, e.g. the LAEs halo filling factor.}
\\
%%%%%%%%%%%%%
%%%%%%%%%%%%%%%%%%% REFEREE %%%%%%%%%%%%%%%%%%%%%%%
%%%%%%%%%%%%% 
%---------------------
%---------------------
%---------------------
\section{Clustering} \label{secClust}
\begin{figure}
\epsscale{.80}
\centering
\includegraphics[width=\columnwidth]{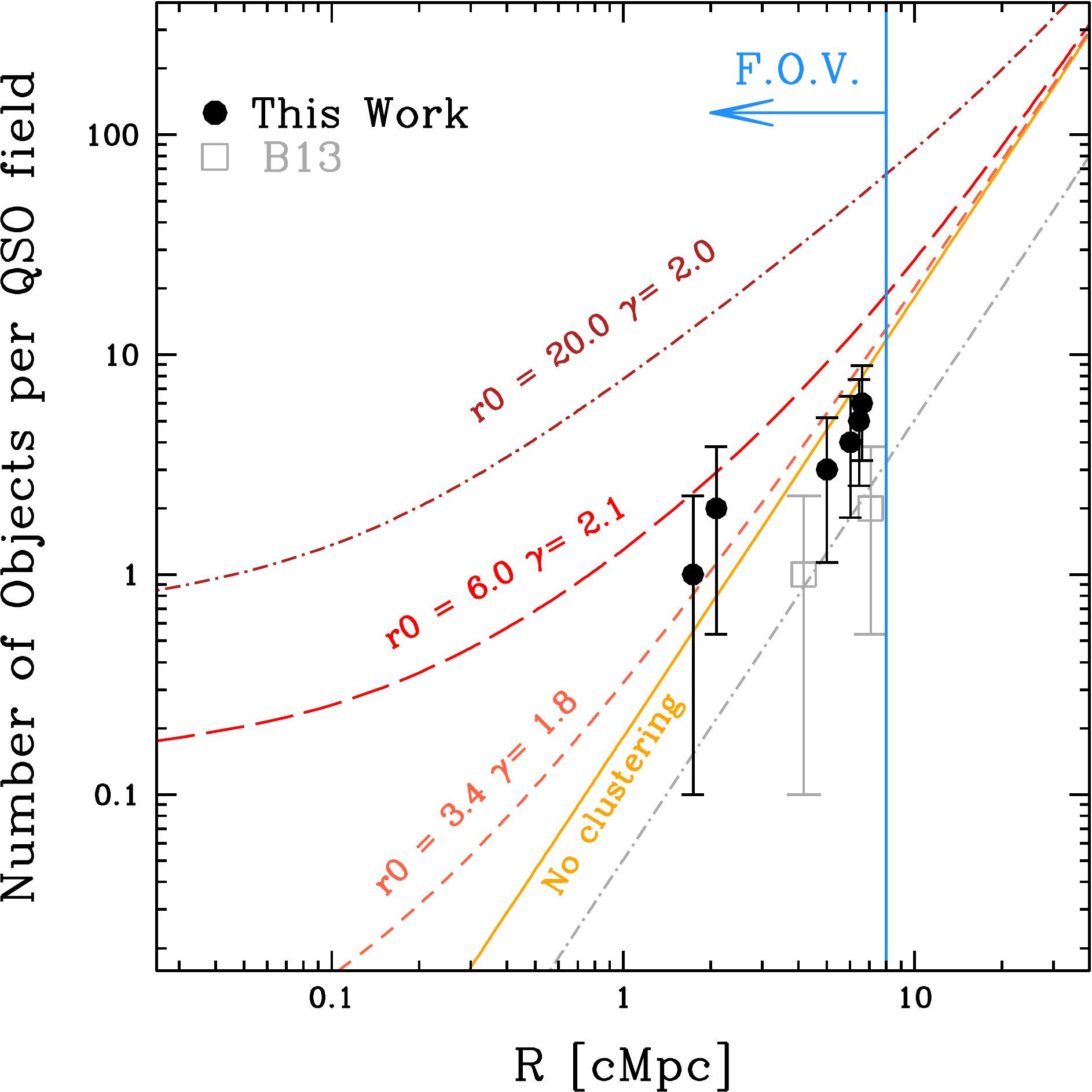}
\caption{Expected number of LAEs, given the depth reached in this study, as a function of projected distance from the quasar, in case of no clustering (e.g. a random distribution of sources, \textit{yellow solid line}), and for some illustrative clustering scenarios, taken from observational studies (\citealt{Zhang13}, \citealt{Ouchi03} and \citealt{McGreer16}, \textit{short-dashed}, \textit{long-dashed} and \textit{dot-short-dashed line}, respectively, see text for details).
%The clustering case as observed at $z\sim$1 (\citealt{Zhang13}, $r0=6$ h$^{-1}$ cMpc, $\gamma$=2.1) and at $3.4<z<5$ are shown with the \textit{red, dashed and light-red dot-dahed line}, respectively.
The counts of LAEs observed in this study are reported (not corrected for completeness). Also, we report the LAEs found by B13, and their expected number of sources in case of no clustering, taking into account the depth of their study (\textit{dot-long-dashed line}). For the sake of clarity, we do not report the respective cases of clustering scenarios. We show the field of view (F.O.V.) encompassed by our study. The errors are the poissonian noise on small counts from \cite{Gehrels86}. The counts of the observed LAEs are consistent with what expected in a blank field.}
\label{figClust}
\end{figure}

%We can estimate the predicted number of galaxies observed at a certain radius with respect to the quasar, depending on the supposed clustering function of quasars.\\
We may also study the environment of PSO J215-16 through a clustering approach.
%if quasars reside in overdensities of galaxies, i.e. quasar and galaxies are clustered, then we expect to detect more sources in the very proximity of the quasar, with respect to what retrieved in a blank field.
If quasars and galaxies are indeed clustered, the excess of probability to find a galaxy at a distance r from a quasar, with respect to a random distribution of sources, can be estimated through the two-point correlation function \citep{DavisPeebles1983}:
%---------------------
\begin{equation}
\xi(\mathrm{r})= (\mathrm{r}/r_{0})^{-\gamma}
\end{equation}
%---------------------
where $r_{0}$ and $\gamma$ are the correlation length and clustering strength, respectively.
%where r$\mathrm{_p}$ and $\rm \pi$ are the projected and line-of-sight galaxy-quasar distance, respectively.
In order to account for redshift distortions on the line of sight, we can consider instead the volume-averaged projected correlation function. This is $\xi(\mathrm{r})$ integrated over a line-of-sight distance $d=2v_{max}/aH(z)$ (with $v_{max}$ maximum velocity from the quasar) and within a radial bin of width [$\mathrm{R_{min}},\mathrm{R_{max}}$] \citep{Hennawi06}:
%---------------------
\begin{equation}
\overline{W}(\mathrm{R_{min},\mathrm{R_{max}}}) =  \frac{{\int^{d/2}_{-d/2}} \int_{\mathrm{R_{min}}} ^{\mathrm{R_{max}}} \xi(\mathrm{r})2\pi dR dr}{V}
\end{equation}
%---------------------
where V is the volume of the cylindrical shell:
%---------------------
\begin{equation}
V = \pi (\mathrm{R_{max}^{2}}-\mathrm{R_{min}^{2}}) d
\end{equation}
%---------------------
Therefore, the number of galaxies that we expect to find around a quasar within a volume V in the presence of clustering is:
%---------------------
\begin{equation} \label{eqclust}
\mathrm{NC} = \mathrm{N} (1+\overline{W}(\mathrm{R_{min},\mathrm{R_{max}}}))
\end{equation}
%---------------------
In case of no clustering, the number of sources expected is:
%---------------------
\begin{equation} \label{eqNOclust}
\mathrm{N} = n V 
\end{equation}
%---------------------
where $n$ is the number density of galaxies per cMpc$^{-3}$, above a certain limit in luminosity. In this scenario, N is equal to the number of galaxies found in a blank field within a cosmological volume V.\\
%integrate this function over $\rm \pi$, obtaining the projected two-point correlation function \citep{DavisPeebles1983}:
%---------------------
%\begin{equation}
%\omega(\mathrm{r_{p}}) = 2 \int_{\mathrm{r_{p}}} ^{\infty} \xi(r) (r^{2}-\mathrm{r_{p}^{2}})^{-1/2} dr 
%\end{equation}
%---------------------
%where $\xi(r)$ can be written as $\xi(r) = (r/r_{0})^{-\gamma}$. We can estimate the number of galaxies in an area of radius r$\rm _{p}$ around a quasar, with:
%---------------------
%\begin{equation} \label{eqclust}
%N = n (1+\omega(\mathrm{r_{p}})) \pi \mathrm{r_{p}}^{2} d
%\end{equation}
%---------------------
In this study, we calculate $n$ by integrating the LAE luminosity function \citep{Ouchi08} down to the luminosity limit obtained by our simulation (1.32$\times$10$^{42}$erg s$^{-1}$, see Section \ref{secSim}). We consider
% $v_{max}=2000$ km s$^{-1}$, equivalent to
a line-of-sight distance given by $v_{max}=4000$ km s$^{-1}$ (consistent with the interval probed by our NB selection, see Sections \ref{secIntro} and \ref{secSim}, and Fig. \ref{figVelDist}), $\mathrm{R_{min}}=0.001$ cMpc and increasing values of $\rm R=\mathrm{R_{max}}$.\\
%Also, other studies of quasar clustering, based on the quasar sample from SDSS-DR7, estimate a strong evolution of the clustering with the redshift: $r_{0}$=21.04 h$^{-1}$ cMpc and $\gamma$=2.0 in the redshift range $3.5<z<5.0$ \citep{Shen09}.
In Fig. \ref{figClust} we show the number of galaxies expected as function of R in case of no clustering (eq. \ref{eqNOclust}), and in different scenarios of quasar-galaxy clustering (eq. \ref{eqclust}). Since there are currently no studies of LAE-quasar clustering at high redshift, we consider for comparison some other illustrative cases. We take values of $r_{0}$ and $\gamma$ obtained by observations of galaxy-quasar clustering at lower-$z$ and LAE-LAE and quasar-quasar clustering at $z\sim5$.
%, obtained varying $r_{0}$ and $\gamma$ within a range of values suggested by observations.
Indeed, studies of galaxy-quasar clustering at $z\sim$1 (\citealt{Zhang13}) estimate $r_{0}$=6 h$^{-1}$ cMpc and $\gamma$=2.1\footnote{We note, however, that the galaxies studied by \cite{Zhang13} are not selected as LAEs but by considering all the sources in the quasar field recovered in the SDSS-Stripe 82 catalog brighter (in the $i$-band) then a certain threshold value, which depends on the field depth. This selection comprehends also passive and red galaxies.}. 
\cite{Ouchi03} derive the clustering properties of LAE at $z=4.86$ from a sample of objects detected in the Subaru Deep Field, for which they obtain $r0=3.35$ h$^{-1}$ cMpc and $\gamma$=1.8.
At high redshift, constraints on the quasar-quasar clustering properties are given by the discovery of a close bright quasar pair, with only 21$\arcsec$ separation, at z$\sim$5 \citep{McGreer16}. The correlation function derived from this pair gives $r_{0}>$20 h$^{-1}$ cMpc and $\gamma$=2.0\footnote{These values are in agreement with the ones found by \citeauthor{Shen07} (\citeyear{Shen07}, $r0$=25.0 h$^{-1}$ cMpc and $\gamma$=2), based on a sample of lower redshift ($z>3.5$) bright SDSS quasars. However, we note that other quasar clustering studies, such as \cite{Eftekharzadeh15}, suggest much smaller clustering scales, with $r0$=7.59 h$^{-1}$ cMpc (obtained from a lower luminosity, z$\sim$3.4, quasar sample; see also Section \ref{secDisc}).}.
%Also, studies of quasar clustering, based on the finding of a close ($r\sim21\arcsec$) quasar pair at z$\sim$5, estimate $r_{0}$=25 h$^{-1}$ cMpc and $\gamma$=2.0 \citep{McGreer15}.
%\footnote{Also, we note that \cite{McGreer15} report the case of quasar-quasar clustering, and not the galaxy-quasar one.}.
We show the number of LAEs found in this study.
We also report the objects recovered by B13 and the number of sources expected in their study in case of no clustering (since their observations are shallower than the ones presented here, with a Ly$\alpha$ luminosity limit of 3.74$\times$10$^{42}$erg s$^{-1}$, the number of background sources expected is lower).
%and from \cite{Shen09}.
Our number counts are consistent with a scenario of no clustering (i.e. the background counts, in line with what obtained in Section \ref{secNumCount}), and do not show evidence of strong clustering in neither of the two quasar fields. 
%and, at r$\rm_{p}\sim$8 cMpc, slightly below the number expected for the z$\sim$1 clustering scenario\footnote{However, we stress that our counts are not corrected for the completeness of our sample.}.
%This result is consistent with a non significant evolution of the galaxy-quasar clustering (and therefore host halo mass) from $z\sim$1 to $z\sim$5.7.
%---------------------

\section{Discussion} \label{secDisc}
We do not find evidence for an overdensity of LAEs in an area of $\sim$37 arcmin$^{2}$ centered on the $z=$5.73 quasar PSO J215$-$16.
Here we investigate possible scenarios to explain our findings.
%-------------
\begin{deluxetable*}{ccccccc}
\tabletypesize{\scriptsize}
\tablecaption{List of z$>$5 quasars whose large-scale fields were inspected for the presence of galaxy overdensities.
We report the quasars names, redshifts, literature references, instruments used and area
covered in each study (in comoving and physical Mpc$^{2}$ at $z\sim$6).
The references are coded as: (1) \cite{Stiavelli05}, (2) \cite{Willott05}, (3)
\cite{Zheng06}, (4) \cite{Kim09}, (5) \cite{Utsumi10}, (6) \cite{Husband13}, (7)
\cite{B13}, (8) \cite{Simpson14}, (9) \cite{Morselli14}, (10) \cite{McGreer14} , (11) this work. 
All these studies, except for \cite{B13} and this work which searched for LAEs,
are based on i-dropout selection.
We report also whether the fields were found overdense (+), underdense (-) or consistent (0) with respect to a comparison blank field.
We note cases in which the same field was found consistent with a blank field when inspected on small scales, while overdense when studied over larger scales (e.g. SDSS J1048+4637 and SDSS J1148+0356). The overdensity reported by \cite{Utsumi10}, even if found with the Subaru SuprimeCam, spreads across an area of $\sim3$Mpc radius, therefore it would have been detected also by searches over smaller fields of view.\label{LitRes}}
\tablewidth{0pt}
\tablehead{
\colhead{Object} & \colhead{Redshift} &  \colhead{Ref} & \colhead{Instrument} & \colhead{FoV} & \colhead{FoV} &  \colhead{Overdensity} \\
   &  &  & &  \colhead{[cMpc$^{2}$]} & \colhead{[pMpc$^{2}$]} &   
}   
\startdata
ULAS J0203+0012 & 5.72 & (7) & VLT FORS2 & 250 & 5.6 & 0\\
SDSS J0338+0021 & 5.03 & (6) & VLT FORS2 & 250 & 5.6 & +\\
SDSS J0836+0054 & 5.82 & (3) & \textit{HST} ACS & 65 & 1.4 & +\\
SDSS J1030+0524 & 6.28 & (1) & \textit{HST} ACS & 65 & 1.4 & + \\
 &  & (2) & GMOS-N & 170 & 3.7 & 0 \\
 &  & (4) & \textit{HST} ACS & 65 & 1.4 & + \\
 &  & (9) & LBT LBC & 3136 & 64.0 & + \\  
SDSS J1048+4637 & 6.20 & (2) & GMOS-N & 170 & 3.7 & 0 \\
 &  & (4) & \textit{HST} ACS & 65 & 1.4 &  0\\
 &  & (9) & LBT LBC & 3136 & 64.0 & + \\
ULAS J1120+0641 & 7.08 & (8) & \textit{HST} ACS & 65 & 1.4 &  0\\
SDSS  J1148+0356 & 6.41 & (2) & GMOS-N & 170 & 3.7 & 0 \\
 &  & (4) & \textit{HST} ACS & 65 & 1.4 & - \\
 &  & (9) & LBT LBC & 3136 & 64.0 & +\\
SDSS J1204$-$0021 & 5.03 & (6) & VLT FORS2 & 250 & 5.6 & +\\
SDSS J1306+0356 & 5.99 & (4) & \textit{HST} ACS & 65 & 1.4 & -\\
SDSS J1411+1217 & 5.95 & (9) & LBT LBC & 3136 & 64.0 & +\\
SDSS J1630+4012 & 6.05 & (4) & \textit{HST} ACS & 65 & 1.4 & +\\
CFHQS J2329$-$0301 & 6.43 & (5) & Subaru SuprimeCam & 4600 & 83.3 & +\\
CFHQS J0050+3445   & 6.25 & (10) & \textit{HST} ACS \& WFC3 & 29 & 0.6 & - \\ 
PSO J215.1512$-$16.0417 & 5.73 & (11) & VLT FORS2 & 206 & 4.5 & 0 
\enddata
\end{deluxetable*}
%-------------------
\begin{itemize}
\item \textit{The overdensity is more extended than our field of view}
\\[1mm]
\cite{Overzier09} and, more recently, \cite{Muldrew15}, through a combination of N-body simulations and semi-analytical models, find that overdensities of galaxies at z$\sim$6 are expected to be very extended, and can cover regions up to $\sim$25$-$30 arcmin radius, corresponding to $\gtrsim$20 pMpc at that redshift.
In the present study we cover only a region of $\sim$1 pMpc transversal radius, and we might be missing a large part of a potential overdensity. We would like to stress that, in the hypothesis that the quasar occupies the center of a $z\sim$6 overdensity similar to the one found by \cite{Toshikawa14} in a blank field, searches in area of $\sim$2$-$3 pMpc should still show evidence of an enhancement in the number of galaxies with respect to a blank field.\\
From an observational perspective, enhancements in the number of galaxies around quasars have been reported on rather modest scales, comparable to ours or even smaller.
However, there are indications, based on LBGs searches, that some quasars are surrounded by overdensities on larger scales, even if a further spectroscopic confirmation is needed (see Section 1).\\
In Table \ref{LitRes} we show a summary of the findings obtained by diverse studies, where they considered different areas and techniques.\\
In our case, in order to discard or confirm this scenario, we would need further observations covering a wider area (e.g with a radius of $\gtrsim$20 arcmin).
\item \textit{The ionizing emission from the quasar is preventing structure formation in its immediate proximities}
\\[1mm]
Strong radiation from a bright quasar can ionize its nearby regions (up to $\sim$1$-$5 pMpc radius around  $z\sim$6 quasars, e.g. \citealt{Venemans15}), with an increase in both the temperature and ionized fraction of the Intergalactic Medium (IGM), and in the intensity of the local UV radiation field.\\
The effects on the visibility of the Ly$\alpha$ radiation in this region are not straightforward to deduce.  
As a consequence of the increase in the UV background radiation field, a higher Ly$\alpha$ transmission flux value is expected around the quasar with respect to the typical IGM environment at the same redshift (\citealt{Bruns12}).
However, in addition to the rise in the UV background, also the nearby IGM temperature increases. Thus, the isothermal virial temperature necessary for gas accretion in the dark matter halo is higher, and the mass needed to form a structure increases (Jeans-mass filtering effect, \citealt{Gnedin00}). Even if the Ly$\alpha$ transmission flux is supposed to be higher, the formation of galaxies itself is suppressed, especially for objects in the low-mass end (e.g. \citealt{Shapiro04}). \cite{Utsumi10} invoke this effect in order to explain the absence of galaxies in a region of $\sim$3 pMpc radius around a z$\sim$6 quasar (see Section 1 and Table \ref{LitRes}). Given the typical sizes of quasar's ionized regions, and that in our study we cover scales of only $\sim$2 pMpc, suppression of galaxy formation due to the quasar ionizing radiation might explain the lack of LAEs.
\item \textit{The bulk of the overdensity population is composed by dusty/obscured galaxies.}
\\[1mm]
Observations find that host galaxies of $z>$5 quasars contain a considerable amount of dust ($\sim$10$^{8}-$10$^{9}$ M$_{\odot}$) and molecular gas ($\sim$10$^{10}$ M$_{\odot}$, see \citealt{Carilli13} for a review). They are already characterized by a metal-enriched medium, comparable to what is observed at low redshift (e.g. \citealt{DeRosa11}). One might foresee that also the galaxies assembling in the proximity of the quasar might be characterized by a high dust/molecular gas content. Indeed, \cite{Yajima15}, implementing a 3D radiation transfer code in a high resolution cosmological simulation, show that overdense regions at $z\sim6$, where quasars are supposedly found, host more evolved, disk-like and massive (M$_{star}\sim$10$^{11}~\mathrm{M}_{\odot}$) galaxies, with respect to an average field at the same redshift. They are characterized by a strong dust extinction (i.e. a low UV radiation escape fraction, f$_{esc}\lesssim$0.1), and a powerful star formation (SFR$\gtrsim$100 M$_{\odot}$ yr$^{-1}$); therefore they are very bright in the IR, with L$\rm _{IR}$ as high as $\sim4\times$10$^{12}~\mathrm{L}_{\odot}$.
These massive and highly obscured objects, whose detection in the UV rest-frame might be hindered by absorption and/or strongly dependent on orientation effects, rather than LAEs (i.e. young, dust-poor star forming galaxies), may be a more suited tracer for high redshift, massive overdensities.\\
Further studies of the environment of high redshift quasars with sub-mm facilities (in particular ALMA, which already successfully detected $z\sim6$ field galaxies, e.g. \citealt{Capak15}, \citealt{Willott15}) would permit to test this scenario, allowing us to recover a possible population of dusty galaxies in the quasar field. However, we note that, due to the small field of view of ALMA (with a size of $\sim20\arcsec$, corresponding to $\rm \sim800~ckpc\sim 110~pkpc$ at $z\sim6$), we would be able to search only the most proximate region around the quasar. A recent study of the fields around three z$>$6.6 quasars with ALMA did not find an excess of dusty galaxies in a region of 65 ckpc radius (\citealt{Venemans16}).
\item \textit{Quasars at high redshift do not inhabit massive dark matter halos}
\\[1mm]
The quasar two-point correlation function at low redshift ($z\lesssim$2.5), as derived from both the 2dF QSO Redshift Survey (\citealt{Croom05}) and the SDSS (\citealt{Ross09}) quasar sample, shows that quasars are commonly associated with average-mass dark matter halos (i.e. M$\rm _{DMH} \sim$ (2-3)$\times$10$^{12}$ M$_{\odot}$), far less massive than the most massive halos at the same redshift ($\sim$10$^{14}-10^{15}$ M$_{\odot}$), independently of the quasar luminosity.\\
At higher-$z$ (3.5$\lesssim z\lesssim$5.4) the scenario is less clear: based on the SDSS sample, \cite{Shen07} calculate an average dark matter host halo mass of (4-6)$\times$10$^{12}$ M$_{\odot}$, slightly higher than the results at lower redshifts. However, more recent studies, based on the final SDSS III-BOSS quasar sample, do not find a  clear evolution of quasar clustering from $z\sim0$ to $z\sim3$ (\citealt{Eftekharzadeh15})\footnote{We note that the quasars considered here are less massive than the ones studied by \cite{Shen07}}.\\
From a theoretical point of view, there have been studies suggesting that quasars also at high redshift (z$\gtrsim$5) inhabit dark matter halos with average masses (i.e. less massive than the most massive halos at that epoch).
In particular, \cite{Fanidakis13} perform simulations based on the semi-analytical model \texttt{GALFORM}, in order to study the relation between quasars and dark matter halos up to $z\sim$6. They show that, in case of models in which AGN feedback is considered, the masses of the dark matter host halos are roughly $\sim10^{12}$ M$_{\odot}$, from $z\sim$0 to $z\sim$6: this is an order of magnitude lower than the most massive halos at $z\sim$6 obtained in the same simulation, and is in agreement with observations at low redshift.
However, as argued by \cite{Simpson14}, it is worth to notice that the simulations by \cite{Fanidakis12} fail to create the most massive black holes (M$\gtrsim$10$^{9}$ M$_{\odot}$) at $z\sim$6, while they appear only at $z\sim$4. Therefore, the claims reported here are to be taken with caution, and may not hold in every scenario.
\end{itemize}

\section{Conclusions}
We studied the environment of the $z$=5.73 quasar PSO J215$-$16 searching for LAEs using broad and narrow-band VLT imaging, on Mpc-scales, i.e. $\sim$2 pMpc $\sim$14 cMpc at the redshift of the quasar. This is the second study in which we do not find evidence of an overdensity of Ly$\alpha$ emitting sources in a quasar field, compared to blank fields (see also B13).
These results may be explained by a variety of different scenarios (i.e. the overdensity is spatially more extended than our field of view, or galaxy formation is prevented by the quasar's ionizing radiation, or the galaxies in the field are mainly dusty, or the quasar is not central to a massive dark matter halo in the early Universe).

Studies on wider areas ($>$20 arcmin radius, corresponding to $\sim$8 pMpc $\sim$ 47 cMpc at the redshift of the quasar), with the support of further, multiwavelength observations (i.e. IR/sub-mm), are required in order to discriminate among these scenarios.

However, it is intriguing to note that overdensities of galaxies around radio-loud sources (both radio-loud galaxies and AGN) have been extensively reported (e.g. \citealt{Venemans07}, \citealt{Wylezalek13}) throughout a wide redshift range (0$<z<$4), and for a couple of cases even at $z\gtrsim$5.2 (\citealt{Venemans04}, \citealt{Zheng06}).
In the future, it appears to be worthwhile to repeat our experiment on $z>6$ radio-loud quasars, whose sample has been substantially increased recently (\citealt{Banados15}), to potentially target the earliest galactic structures.
 
\acknowledgments
Based on observations made with ESO Telescope at the La Silla Paranal Observatory, under program ID 091.A-0677(A).\\
We thank the referee for their careful reading of this manuscript and their constructive and helpful comments and suggestions.\\
B.P.V, E.P.F. and F.W. acknowledge funding through the ERC grant "Cosmic Dawn". Support for R.D. was provided by the DFG priority program 1573 "The physics of the interstellar medium". Support for R.O. was provided by CNPq programs 459040/2014-6 and 400738/2014-7. C.M. thanks the IMPRS for Astronomy \& Cosmic Physics at the University of Heidelberg. 
C.M. thanks N. Fanidakis, L. Morselli, C. Garcia, R. Gilli and M. Onoue for useful discussion on this project.

{\it Facilities:} \facility{VLT:Antu(FORS2)}.

\appendix

\section{Star Formation Rate Estimates of LAE candidates}  \label{secSFR} \label{AppendixA}
We infer here a star formation rate (SFR) estimate for our possible LAE candidates. We use as SFR tracer the luminosity of the Ly$\alpha$ emission line.

We obtain the integrated luminosity of the Ly$\alpha$ line from the flux density observed in the narrow band filter ($ f_{\mathrm{NB}}$):
%---------------------
\begin{equation}
 L_{\mathrm{Ly\alpha}}= f_{\mathrm{NB}} 4 \pi d_{L}^{2}  \Delta \nu_{\mathrm{NB}}
\end{equation}
%---------------------
where $d_{L}$ is the luminosity distance at the redshift of the quasar and $\Delta \nu_{\mathrm{NB}}$ the width of the narrow band filter.\\
In this estimate, we do not correct for the contribution of the continuum, expected to be very faint (none of our LAE candidates are detected in the broad bands).\\
From $L_{\mathrm{Ly\alpha}}$ we can derive the luminosity of the H$\alpha$ emission line ($L_{\mathrm{H\alpha}}$). Assuming the case-B recombination (\citealt{Osterbrock89}), the conversion is given by $L_{\mathrm{Ly\alpha}} = 8.7 \times L_{\mathrm{LH\alpha}}$. Then, we use the following relation between SFR and L$\rm _{H\alpha}$ \citep{Kennicut12}:
%---------------------
\begin{equation}
\log \frac{\mathrm{SFR}_{\mathrm{Ly\alpha}}}{\mathrm{M_{\odot} \quad yr^{-1}}}= \log \frac{L_{\mathrm{H\alpha}}}{\mathrm{erg \quad s^{-1}}} - 41.27
 \label{eqSFR}
\end{equation}
%--------------------- 
We obtain SFR estimates in the range between (1.2$\pm$0.2)$-$(3.2$\pm$0.3) M$_{\odot}$ yr$^{-1}$ (all the SFR values, together with the respective L$_{Ly\alpha}$, are reported in Table \ref{tab1})
%\textbf{Due to the numerous assumptions adopted while estimating the SFRs, the values reported here can be considered only as lower limits.} 
%---------------------
%-------------------
%%%%%%%%%%%%%
%%%%%%%%%%%%%%%%%%% REFEREE %%%%%%%%%%%%%%%%%%%%%%%
%%%%%%%%%%%%%
%---------------------
In all our analysis, we do not consider possible absorption due to the galactic dust.
Even if LAEs are thought to be rather dust-poor objects (e.g. \citealt{Garel15}), there have been evidence for a non negligible fraction of dusty LAEs (\citealt{Pentericci09}). The interstellar neutral gas, its geometry and dynamic, gives also an important contribution to the effective Ly$\alpha$ photon escape fraction. Indeed a higher $L_{\mathrm{H\alpha}}/L_{\mathrm{Ly\alpha}}$ ratio is expected in case of a lower Ly$\alpha$ photon escape fraction. Finally, we neglect the effect of the significantly neutral intergalactic medium at the high redshift under consideration.\\
All these contributions concur in reducing the estimated SFRs: the values reported 
here can thus be considered only as lower limits.
%%%%%%%%%%%%%
%%%%%%%%%%%%%%%%%%% REFEREE %%%%%%%%%%%%%%%%%%%%%%%
%%%%%%%%%%%%%
%--------------------
%---------------------
\section{Lyman Break Galaxies Analysis}
\begin{figure}
\epsscale{.80}
\begin{center}
\includegraphics[width=0.3\columnwidth]{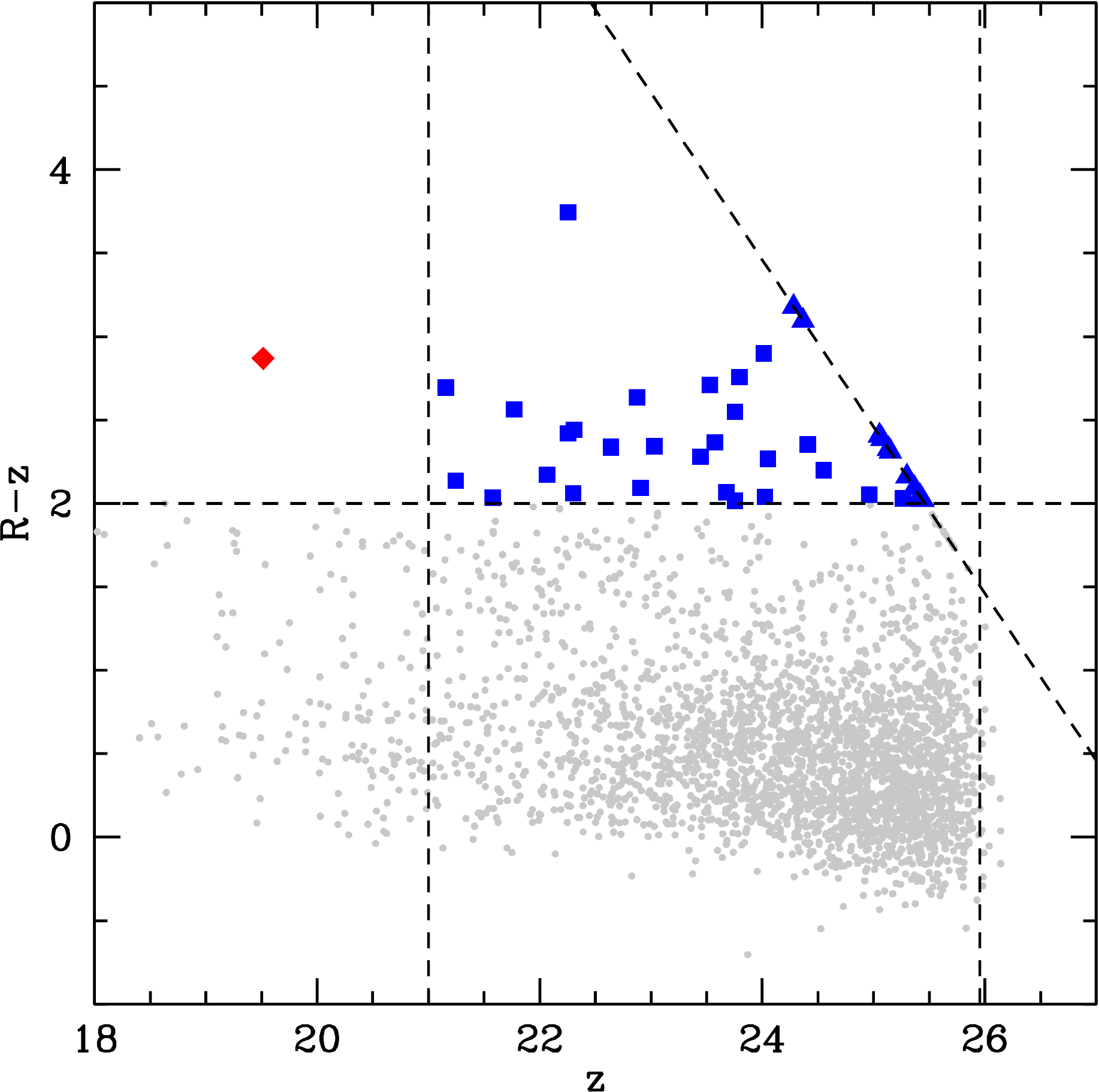}
\caption{Color-magnitude diagram (R-z) vs z. In \textit{grey points} all the sources detected in our field, considering the z frame as reference. The LBG candidates found from the z frame (37) are reported in \textit{blue squares}. We show with \textit{vertical dashed lines} the lower limit on the z magnitude (z $>$ 21) and the 5$\sigma$ upper limit (z=25.96). The \textit{horizontal line} highlights our color criteria. The \textit{diagonal dashed line} displays the 2$\sigma$ limit magnitude in the R frame. The objects not detected in the R frame at 2$\sigma$ level are reported in \textit{blue triangles}. The \textit{red diamond} shows the quasar position.}
\label{figColorMag}
\end{center}
\end{figure}

In addition to the LAE selection, we also search for LBGs using the dropout technique. Since LBGs are expected to be characterized by a strong UV continuum, observed in the z filter,  we use the catalog obtained taking the z frame as our reference image. We consider all the sources with S/N$>$4 in z, and we apply only a selection using the broad band filters: we ask for a red R-z color (R-z$>$2) and, since we expect galaxies at these redshift to be faint, we require z$>$21. \\
We recover 37 LBG candidates: we report the color-magnitude (R-z) vs z in Fig. \ref{figColorMag}.
It is worth to notice that, in the selection of LBG candidates, we are mainly limited by the depth of the R image (R$_{2\sigma}=27.46$) rather than by the z one; indeed, the faintest sources with (R$_{2\sigma}-$z)$>$2 would have z$\leq$25.46 in our analysis (z$\rm _{5\sigma, lim}=25.96$).
%-------------------
%---------------------

For comparison, we refer to the works by \cite{Brammer12} and \cite{Skelton14}, who compiled catalogs for some well-known extra-galactic fields, completed with spectroscopic and photometric information.
We consider four fields, that can be used as comparison blank fields for our study: All-wavelength Extended Groth Strip International Survey (AEGIS), the Cosmic Evolution Survey (COSMOS), the Great Observatories Origins Survey Northen field (GOODS-N) and the UKIRT InfraRed Deep Sky Survey (UKIDSS) Ultra Deep Field (UDS).

\begin{figure}[h]
\epsscale{.80}
\centering
\includegraphics[width=0.5\columnwidth]{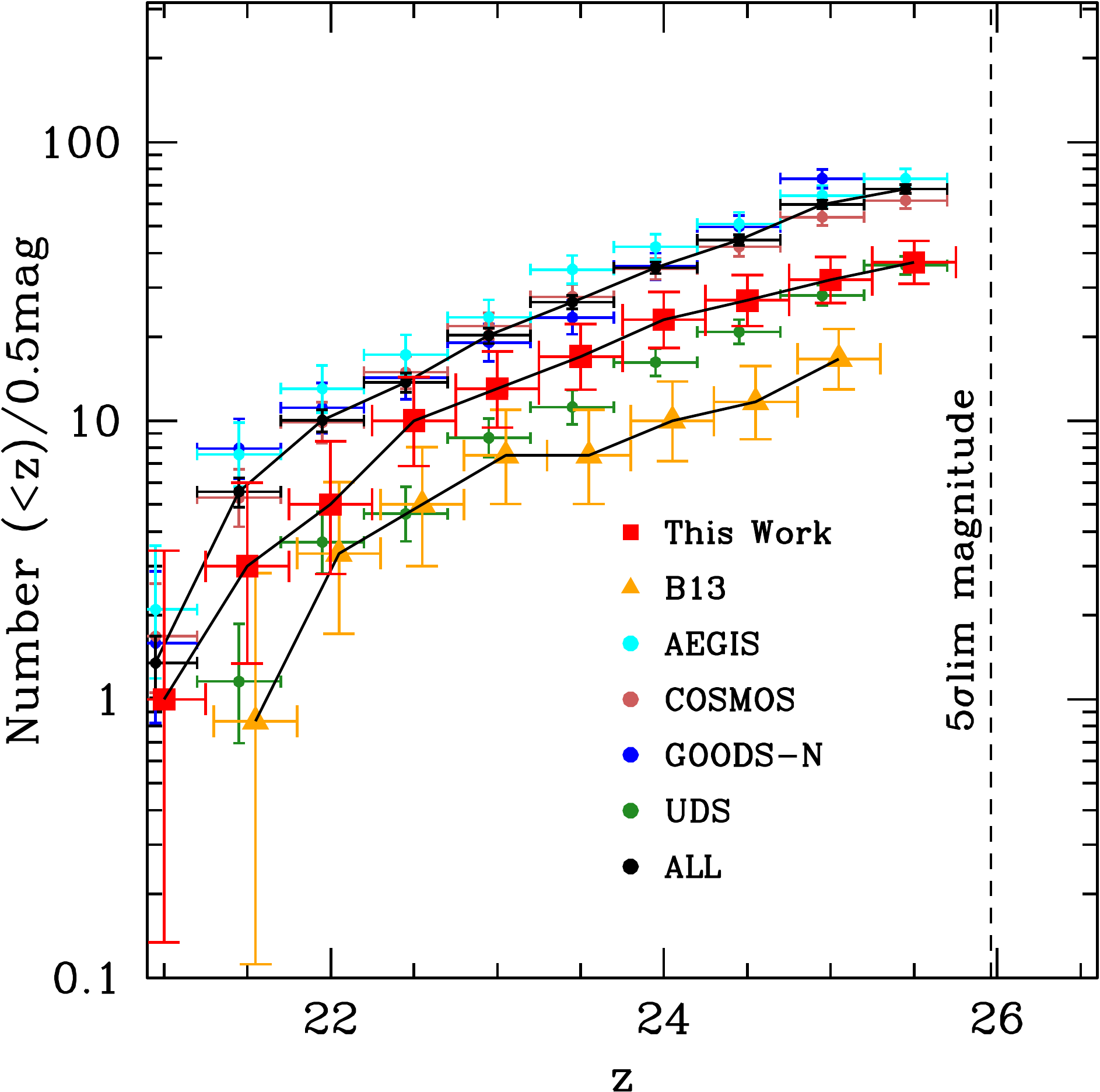}
\caption{Cumulative Number Counts of LBGs at z$\sim$6. All the counts are scaled for the effective area in our study. We report the candidates found around the $z$=5.73 quasar PSO J215$-$16 (\textit{red squares}) and the quasar ULAS J0203+0012 (\textit{orange triangles}, B13), slightly shifted with respect to the other fields in order to avoid confusion. We show the sources selected, using the same selection criteria and image depth, in four comparison blank fields (AEGIS-\textit{cyan}, COSMOS-\textit{brown}, GOODS-N-\textit{blue} and UDS-\textit{green circles}). The result obtained considering all the blank fields together is shown with \textit{black circles}. The errors are taken from the poisson noise in case of low counts statistics (\citealt{Gehrels86}). }
\label{figNumbCountLBG}
\end{figure}

We selected LBG candidates from the catalogs imposing the same selection criteria as in the present study. In order to account for the different image depths, we consider only sources with z and R magnitude lower than the limits in our field (z$\rm < z_{5\sigma,lim}=$ 25.96 and R$\rm < R_{2\sigma,lim}=$ 27.46). The GOODS-N and UDS fields are shallower than our images in both R and z and only in the z frame, respectively. In these cases, we take as limits the corresponding values provided by \cite{Skelton14} ($\rm z_{5\sigma,lim, GOODSN}= 25.5$, $\rm R_{2\sigma,lim, GOODSN}= 27.19$ and $\rm z_{5\sigma,lim, UDS}= 25.9$).\\
We also consider the sample of LBG candidates recovered by B13 around ULAS J0203+0012 (20 sources), where they used an analogous selection method as the one described here. They reach limit magnitudes in the z and R bands of $\rm z_{5\sigma, lim, B13}= 25.14$ and $\rm R_{2\sigma, lim, B13}=27.29$.\\
%---------------------
\begin{small}
\begin{deluxetable*}{cccccccccc}[h]
\tabletypesize{\scriptsize}
\centering
\tablecaption{Field names, coordinates, effective areas and technical characteristics for the R and z filters of our comparison fields and the one studied here. The areas analyzed in this study and in B13 differ due to the diverse masking.
References from the literature are: (1) \cite{Hildebrandt09}, (2) \cite{Erben09}, (3) \cite{Capak04}, (4) \cite{Furusawa08}, (5) \cite{B13}, (6) This work.}
\tablewidth{0pt}
\tablehead{
\colhead{Field} & \colhead{RA} &  \colhead{DEC} & \colhead{Effective Area} & \colhead{$\rm \lambda_{c, R}$} & \colhead{$\rm \Delta \lambda_{R}$} & \colhead{$\rm \lambda_{c, z}$} & \colhead{$\rm \Delta \lambda_{z}$}  & \colhead{Instrument} & \colhead{Reference}\\
   & \colhead{(J2000.0)} & \colhead{(J2000.0)} & \colhead{[arcmin$^{2}$]} & \colhead{[$\rm\AA$]} & \colhead{[$\rm\AA$]} & \colhead{[$\rm\AA$]} & \colhead{[$\rm\AA$]} & 
}   
\startdata
AEGIS   & 14:18:36.00 & +52:39:0.00 & 88 & 6245 & 1232 & 8872 & 1719 & MegaCam@CFHT & (1), (2)\\
COSMOS  & 10:00:31.00 & +02:24:0.00 & 154 & 6245 & 1232 & 8872 & 1719 & MegaCam@CFHT & (1), (2)\\
GOODS-N & 12:35:54.98 & +62:11:51.3 & 93 & 6276 & 1379 & 9028 & 1411 & Suprime-Cam@Subaru & (3)\\
UDS     & 02:17:49.00 & −05:12:2.00 & 192 & 6508 & 1194 & 9060 & 1402 & Suprime-Cam@Subaru & (4)\\
B13     & 02:03:32.38 & 00:12:29.06 & 44 & 6550 & 1650 & 9100 & 1305  & FORS2@VLT & (5)\\
This work & 14:20:36.39 & -16:02:29.94 & 37 & 6550 & 1650 & 9100 & 1305 & FORS2@VLT & (6)
\enddata
\label{tabFields}
\end{deluxetable*}
\end{small}
%-------------------
In Table \ref{tabFields} we report information on the comparison fields and on our field, i.e. coordinates, effective areas, literature references and characteristics of R and z filters. Although the different fields were imaged with slightly different filter sets, the redshift windows covered are large ($\mathrm{\Delta} z \sim 1.2$) and corresponding to the one spanned in the present study\footnote{With the filters used, in the AEGIS and COSMOS fields we span $ 5.2 \lesssim z\lesssim 6.3$, $ \mathrm{\Delta} z\sim1.1$, in the GOODS-N field $ 5.2 \lesssim z\lesssim 6.4$, $ \mathrm{\Delta} z\sim1.2$, while in the UDS field $ 5.4 \lesssim z\lesssim 6.5$, $\mathrm{\Delta} z\sim1.1$. In the present study and in B13 we are selecting sources in $ 5.2 \lesssim z\lesssim 6.5$, $ \mathrm{\Delta} z\sim1.3$.}.
We report the cumulative number counts, scaled to our effective area, of the sources found in the four blank fields and around the quasars (Fig. \ref{figNumbCountLBG}).
The difference between the counts obtained in the UDS field with respect to the counts in the other blank fields may be due to a diverse contribution of contaminant sources. Indeed, the UDS field was imaged through a R filter slightly redder than the ones used in the other fields (see Table \ref{tabFields}): this might turn into a more conservative selection of high-redshift LBGs. Recent studies suggest also that the UDS field might be intrinsically underdense in z$\sim$6 galaxies with respect to other well-known fields (\citealt{Bowler15}, \citealt{Bouwens15}).\\
We can compare the cumulative number counts of LBGs in the field of the quasar studied here with the ones found in the comparison blank fields. In order to avoid incompleteness issues in the low-luminosity end, we take only sources with R$<$R$\rm _{lim,5\sigma}$, considering the GOODSN field, which is the shallowest among our fields (R$\rm _{lim,5\sigma}$=26.2). Taking into account our color selection criterion, we obtain a resulting z magnitude limit of 24.2. At this limit, the counts of the LBG candidates in our quasar field is consistent within 1$\sigma$ with the counts in the UDS field, and lower than the ones in AEGIS, COSMOS and GOODS-N fields by $\sim$1.7, 1.3 and 1.1$\sigma$ respectively.
The quasar field analyzed here appears only marginally ($\sim$1.1$\sigma$) denser than the one studied in B13.
%if we assume the same limit magnitude (z$<$z$\rm _{5\sigma,lim,B13}=25.14$)
These results are hence in agreement with B13, where no overdensity of LBGs with respect to a blank field was found.

Some general caveats are to be taken into account. Considering our broad selection criteria, both our sample and the ones derived from the comparison fields might be contaminated by red, lower redshift sources.
We employ the further information provided in the catalogs of the comparison fields in order to better characterize the sources retrieved by our LBGs selection. We can consider the available photometric redshift estimates, computed with the public code EAZY (\citealt{Brammer08}), which take into account all the photometric information present in the catalog.
We take only the objects with a reliable redshift estimate, as based on the quality parameter $Q_{z}$ ($Q_{z}<2.0$, see \citealt{Brammer08}), and for which z$_{phot}\geq$5.0. Only the 6\%, 10\%, 4\% and 9\% of the LBG sample from, respectively, AEGIS, COSMOS, GOODS-N and UDS field could be identified as high-redshift galaxies (see Table \ref{tabFieldsCounts}), while the vast majority was better fitted by a z$\sim$1 galaxy model.
This simple test shows how the selection criteria used here, without the help of further bands, lead us to a highly contaminated sample.

In summary the LBGs selection also does not reveal a possible overdensity around the quasar.
However, due to the wide redshift range considered, an enhancement in the number of LBGs in the quasar field would represent an indication, more than solid evidence, for the presence of an overdensity of galaxies in the proximity of the quasar.
%---------------------
\begin{small}
\begin{deluxetable}{ccc}[h]
\tabletypesize{\scriptsize}
\centering
\tablecaption{Field names, total number of LBG candidates retrieved by our photometric cuts
and number of sources with photometric redshift estimates
corresponding to $z_{phot}\ge5$.}
\tablewidth{0pt}
\tablehead{
\colhead{Field} & \colhead{Number LBGs} &  \colhead{Number phot LBGs}
}   
\startdata
AEGIS   & 176 & 11\\
COSMOS  & 257 & 26\\
GOODS-N & 186 & 7\\
UDS     & 187 & 17
\enddata
\label{tabFieldsCounts}
\end{deluxetable}
\end{small}
%-------------------
%---------------------
%---------------------
\pagebreak
%---------------------
%---------------------

\clearpage

\end{document}